\documentclass[aps,prl,reprint,amsfonts,twocolumn,floatfix,hidelinks,superscriptaddress,nofootinbib]{revtex4-2}
\usepackage{graphicx}
\usepackage{dcolumn}
\usepackage{bm}
\usepackage[table]{xcolor}
\usepackage{float}
\usepackage{soul}
\usepackage{hyperref}
\hypersetup{colorlinks=true,citecolor=blue,linkcolor=blue,urlcolor=blue,pdfstartview=FitH}
\usepackage{braket}
\usepackage{array}
\usepackage{colortbl}
\usepackage{booktabs}
\usepackage{amsmath}
\usepackage{verbatim}  
\usepackage{tikz}
\renewcommand{\arraystretch}{1.4} 
\definecolor{headercolor}{rgb}{0.8, 0.8, 0.8}

\definecolor{ao(english)}{rgb}{0.0, 0.5, 0.0}

\definecolor{midnightblue}{rgb}{0.39,0.58,0.93}

\usepackage{tabularx}
\newcolumntype{Y}{>{\centering\arraybackslash}X}
\newcolumntype{Z}{>{\centering\arraybackslash}p{0.25\columnwidth}} 

\begin{document}

\preprint{APS/123-QED}

\title{Practical roadmap to measurement-altered criticality in Rydberg arrays}

\author{Stephen Naus}
\email{snaus@caltech.edu}
\affiliation{
Department of Physics and Institute for Quantum Information and Matter, \\ California Institute of Technology, Pasadena, California 91125, USA}

\author{Yue Liu}
\affiliation{
Department of Physics and Institute for Quantum Information and Matter, \\ California Institute of Technology, Pasadena, California 91125, USA}

\author{Sara Murciano}
\affiliation{Universit\'e Paris-Saclay, CNRS, LPTMS, 91405, Orsay, France}

\author{Pablo Sala}
\affiliation{
Department of Physics and Institute for Quantum Information and Matter, \\ California Institute of Technology, Pasadena, California 91125, USA}
\affiliation{Walter Burke Institute for Theoretical Physics, California Institute of Technology,\\ Pasadena, California 91125, USA}
\author{Manuel Endres}
\affiliation{
Department of Physics and Institute for Quantum Information and Matter, \\ California Institute of Technology, Pasadena, California 91125, USA}
\author{Jason Alicea}
\affiliation{
Department of Physics and Institute for Quantum Information and Matter, \\ California Institute of Technology, Pasadena, California 91125, USA}
\affiliation{Walter Burke Institute for Theoretical Physics, California Institute of Technology,\\ Pasadena, California 91125, USA}

\date{\today}

\begin{abstract}
Weak measurements have been predicted to dramatically alter universal properties of quantum critical wavefunctions, though experimental validation remains an open problem.  Here we devise a practical scheme for realizing measurement-altered criticality in a chain of Rydberg atoms tuned to Ising and tricritical Ising phase transitions. 
In particular, we show that projectively measuring a periodic subset of atoms alters quantum critical correlations in distinct ways that one can control via the choice of measured sites and the measurement outcomes.  While our protocol relies on post-selection, the measurement outcomes yielding the most dramatic consequences occur with surprisingly large probabilities: $O(10\%)$ with chains featuring $O(100)$ sites. Characterizing the proposed post-measurement states requires only an adjustment in the post-process averaging of outcomes used to characterize unmeasured critical states, resulting in minimal additional experimental overhead for demonstrating measurement-altered criticality.

\end{abstract}

\maketitle

\textbf{\emph{Introduction.}}~Learning about quantum matter through measurement non-unitarily alters the underlying wavefunction, thereby furnishing a control knob complementary to conventional unitary evolution.  Quantum critical states, describing systems tuned to a phase transition, provide an attractive setting for developing this paradigm.  Criticality engenders universal properties that, by construction, display exquisite sensitivity to perturbations---including from measurement.  Indeed, gaining even arbitrarily small amounts of information about a quantum critical wavefunction can qualitatively alter its  correlations and entanglement in a manner dependent on the measurement basis and outcome \cite{AltmanMeasurementLL}.  This phenomenon of `measurement-altered quantum criticality' has by now been theoretically explored in numerous settings \cite{AltmanMeasurementLL, sun2023,Ashida23,EhudMeasurementIsing,JianMeasurementIsing,usmeasurementaltered,Paviglianiti2023,Ludwig2024HighlyComplex,Lee2023,tang2024,Yue_gapless2024}. 
Practical implications range from understanding decoherence effects \cite{Lee2023,Zou2023,Ashida23} to optimizing against errors in teleportation protocols \cite{Qteleport} (see also Ref.~\onlinecite{GuoYiTeleportation}). 

\begin{figure}[ht!]
\includegraphics[width=\linewidth]{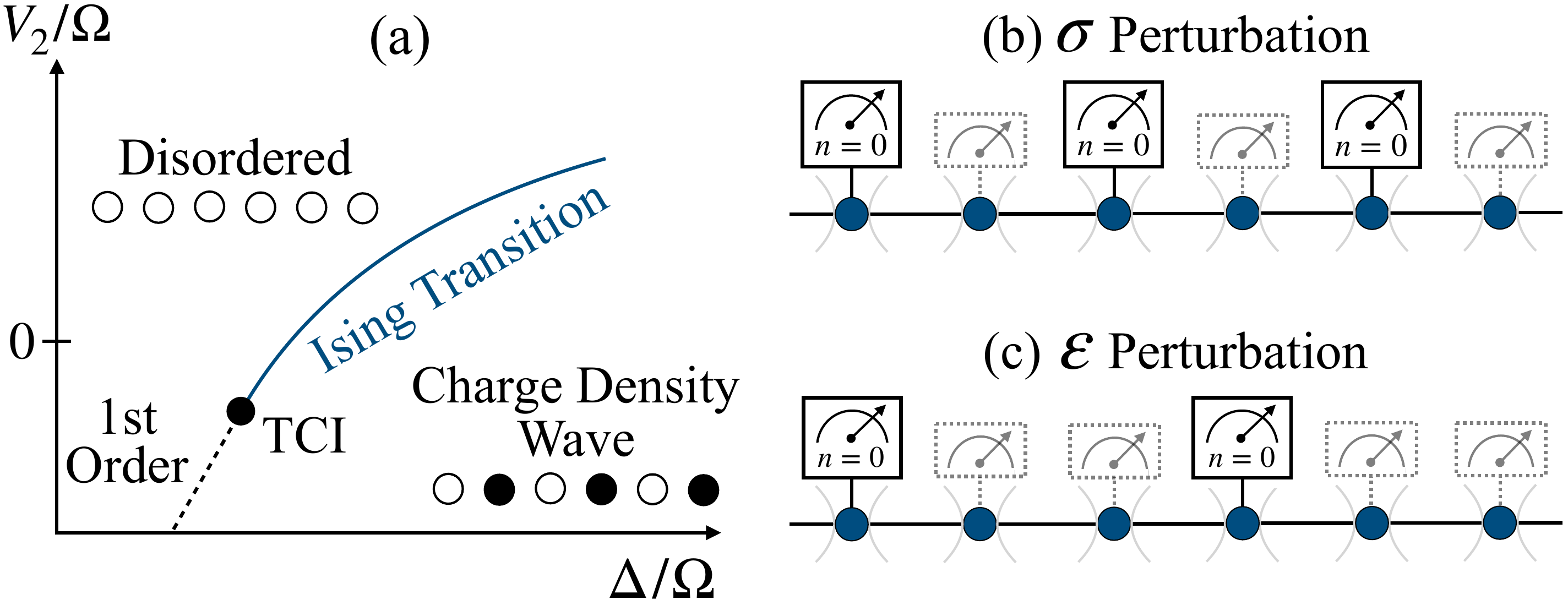}
\caption{{\bf Phase diagram and measurement protocols.} (a) Schematic phase diagram of the blockaded Rydberg chain Hamiltonian, Eq.~\eqref{eq:model}, featuring charge density wave and disordered states separated by an Ising transition line terminating at a tricritical Ising (TCI) point. (b,c) Example protocols for measurement-altered criticality associated with (b) $\sigma$ and (c) $\varepsilon$ CFT perturbations.  Modified correlations follow by simply measuring all sites simultaneously, and averaging over outcomes subject to the post-selection conditions indicated in the solid measurement icons.} 
\label{fig:Setup}
\end{figure}

Experimental exploration of measurement-altered criticality requires a versatile platform capable of initializing a critical ground state with high fidelity, measuring to appropriately modify the wavefunction, and detecting the resulting changes in universal properties.  Wavefunction modifications are commonly studied theoretically using weak measurements \cite{ sun2023,Ashida23,EhudMeasurementIsing,JianMeasurementIsing,Paviglianiti2023,Ludwig2024HighlyComplex,Lee2023,tang2024,Yue_gapless2024}, e.g., mediated by ancillae that are unitarily entangled with the critical degrees of freedom and then projectively measured \cite{AltmanMeasurementLL,usmeasurementaltered}.
Weak measurements can modestly restructure the amplitudes in the critical wavefunction, facilitating field-theoretic predictions for the impact on long-distance behavior
\cite{AltmanMeasurementLL,usmeasurementaltered,Ashida23,sun2023,EhudMeasurementIsing,JianMeasurementIsing,Yue_gapless2024,Ludwig2024HighlyComplex,Lee2023,tang2024,Paviglianiti2023,Hoshino2024,Nahum2025,Putz2025}. 
Verifying these predictions poses an intrinsic problem, however, since measurements restructure the wavefunction in a manner dependent on the information gleaned---which is inherently probabilistic.  
Revealing measurement-altered criticality 
requires either post-selecting for specific target outcomes---whose probability generically decreases exponentially with system size---or non-standard averaging procedures \cite{AltmanMeasurementLL, usmeasurementaltered, Garratt23,Ludwig2024HighlyComplex, EhudMeasurementIsing}.

We provide a practical roadmap for addressing these challenges using a single chain of laser-excited Rydberg atoms trapped in optical tweezers, which can realize Ising and tricritical Ising quantum critical wavefunctions \cite{FSS,Bernien_2017,Keesling_2019,Slagle2021,Scholl_2023,Fang2024expcrit} (among other universality classes \cite{Whitsitt2018,Chepiga2019,Chepiga2021,Z3Rydberg_Sachdev,Chepiga_2024,zhang2025probing,Chepiga2025}) via adiabatic state preparation; see Fig.~\ref{fig:Setup}(a).   We develop an ancilla-free protocol that is useful to conceptually divide into two stages.  First, projectively measure the state of the atoms at a \emph{subset} of sites.

\begin{figure}
\begin{tikzpicture}
  \node[anchor=south west,inner sep=0] (image) at (0,0) {\includegraphics[width=\linewidth]{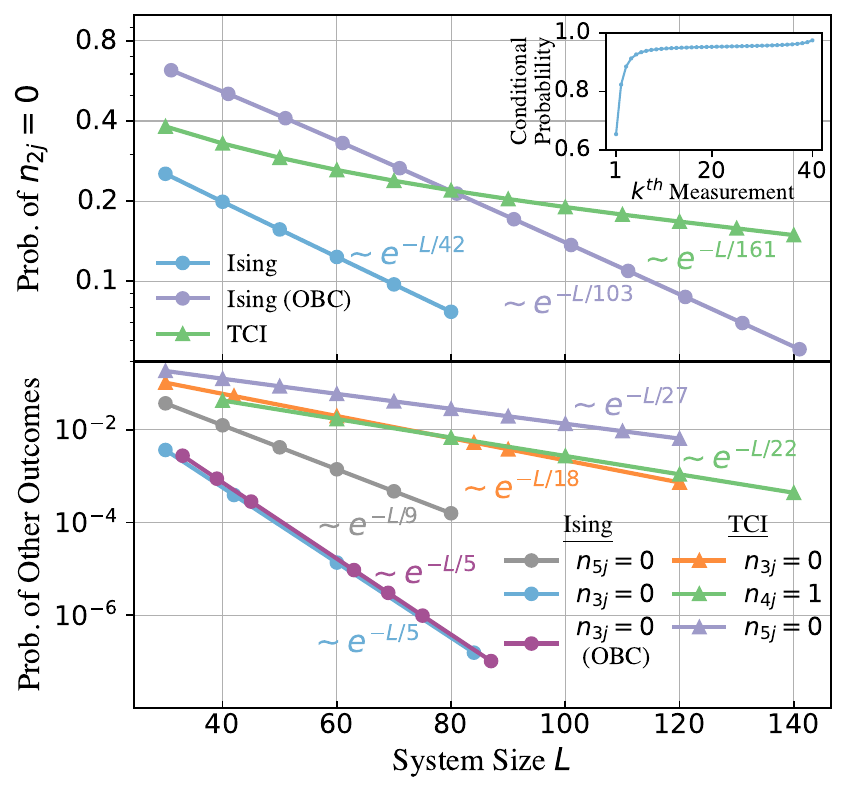}};
  \begin{scope}[x={(image.south east)},y={(image.north west)}]
    \node at (0.09,1.0) {\textbf{(a)}};
    \node at (0.09,.54) {\textbf{(b)}};
  \end{scope}
\end{tikzpicture}
 \caption{\textbf{Enhanced probability outcomes.} Post-selection probabilities for (a) $\{n_{2j} = 0\}$ and (b) other outcomes in critical Ising and TCI Rydberg chains. \emph{Panel (a) inset:} conditional probability of measuring $n_{2k} = 0$, given prior $n_{2(j<k)} = 0$ outcomes for an $L = 80$ critical Ising chain with periodic boundaries.} 
\label{fig:PSprobs}
\end{figure}

Judiciously choosing both the pattern of measured sites and post-selected measurement outcomes, e.g., as indicated by the solid measurement icons in Fig.~\ref{fig:Setup}(b,c), controls universal modifications to critical correlations.  These modifications  manifest with either open or periodic boundary conditions 
and can be probed via subsequently measuring the remainder of the chain [faded measurement icons in Fig.~\ref{fig:Setup}(b,c)].  We show that the post-selection probabilities for the target measurement outcomes yielding the most dramatic effects are surprisingly high for experimentally relevant system sizes; see Fig.~\ref{fig:PSprobs}.  More practically, the same measurement-altered correlations emerge upon projectively measuring \emph{all} sites simultaneously and simply averaging over a restricted set of measurement outcomes (instead of all outcomes, which would return correlations from the unmeasured theory).  Our scheme thus entails minimal overhead beyond that required to probe pristine quantum critical correlations, facilitating near-term experimental realization.

\textbf{\emph{Model and `pure' criticality.}}~We model a length-$L$ Rydberg chain with the minimal Hamiltonian \cite{FSS} 
\begin{equation}
\hat{H} = 
\sum_j\left[\frac{\Omega}{2}(\hat{b}_j+\hat{b}_j^\dagger)-\Delta \hat{n}_j + V_1\hat{n}_j\hat{n}_{j+1} + V_2 \hat{n}_j\hat{n}_{j+2}\right],
\label{eq:model}
\end{equation}
where $\hat{b}_j$ is a hard-core boson operator and $\hat{n}_j = \hat{b}_j^\dagger \hat{b}_j$ with eigenvalues $0$ (ground state) and $1$ (Rydberg excited state).  The first two terms encode, within the rotating wave approximation, driving of the atoms with Rabi frequency $\Omega$ and detuning $\Delta$ from resonance; $V_{1,2}$ represent first- and second-neighbor interactions mediated by the Rydberg excited states.  We assume that $V_1>0$ dominates all couplings, energetically imposing the blockade constraint that no two neighboring atoms can simultaneously enter the Rydberg state.  Within the blockaded subspace, one can effectively realize either sign of $V_2$ in experiment \cite{Slagle2021,Scholl_2023}.   Unless specified otherwise we consider an even-length, periodic chain for which $\hat{H}$ preserves both translation $T_x$ and bond-reflection $R_x$ symmetries.  

The phase diagram of $\hat{H}$ [Fig.~\ref{fig:Setup}(a)] features a trivial symmetric phase and a two-fold degenerate CDW phase wherein Rydberg excitations occur predominantly on every other site, doubling the unit cell and spontaneously breaking $T_x$ and $R_x$.  Starting from $V_2>0$, these phases are separated by a continuous Ising transition that, on decreasing $V_2$, terminates at an integrable tricritical Ising (TCI) point \cite{FSS} and then becomes first-order. Throughout we consider a Rydberg chain prepared into the ground state $\ket{\psi_c}$ of either the Ising or TCI transitions, which are respectively governed by conformal field theories (CFTs) with central charge $c = 1/2$ and $7/10$.

Symmetries allow one to relate microscopic operators to CFT fields that govern their leading long-distance correlations \cite{Slagle2021}.  Lattice operators that transform trivially under $T_x$ and $R_x$ generically map to the `thermal perturbation' $\varepsilon$ that one must fine-tune away to reach criticality, e.g.,
\begin{equation}
    \hat{\varepsilon}_{j+1/2} \equiv (\hat{n}_j + \hat{n}_{j+1})/2 \sim c_1 + c_2 \varepsilon + \cdots
    \label{eq:epsilon_map}
\end{equation}
with non-universal $c_{1,2}$ and subleading terms indicated by the ellipsis.  Equation~\eqref{eq:epsilon_map} implies that connected two-point correlations exhibit universal power-law decay: $\langle \hat{\varepsilon}_{j+1/2} \hat{\varepsilon}_{j'+1/2}\rangle_c \sim \langle \varepsilon(x)\varepsilon(x') \rangle_c \sim |j-j'|^{-2\Delta_\varepsilon}$ 
with scaling dimension $\Delta_{\varepsilon} = 1$ (Ising) or $1/5$ (TCI) dictated by the underlying CFT.  

Moreover, the microscopic CDW order parameter maps to the CFT `spin field' $\sigma$,
\begin{equation}
     \hat{\sigma}_{j+1/2} \equiv (-1)^j (\hat{n}_j - \hat{n}_{j+1})/2 \sim \sigma + \cdots.
     \label{eq:sigma_map}
\end{equation}
Both $\hat{\sigma}_{j+1/2}$ and $\sigma$ acquire a minus sign under $T_x$ and $R_x$, and hence symmetry dictates that $\langle \hat{\sigma}_{j+1/2}\rangle = \langle \sigma \rangle = 0$.  Two-point order-parameter correlations, however, also decay as a power-law, but with scaling dimension $\Delta_\sigma = 1/8$ (Ising) or $\Delta_\sigma = 3/40$ (TCI).  Chains with open boundaries explicitly break $T_x$, seeding CDW order at the edges \cite{Slagle2021}. 
Consequently, even in an otherwise uniform critical chain, $\langle \hat{\sigma}_{j+1/2}\rangle$ and $\langle \sigma\rangle$ become finite and decay towards zero into the bulk with a functional form dependent on $\Delta_\sigma$; see Appendix~\ref{app:A} for details.  Open chains thereby enable extraction of scaling dimensions via relatively simple one-point expectation values.  

{\bf \emph{Measurement-altered criticality protocol.}}~We examine changes in universal
properties of the critical state $\ket{\psi_c}$ after measuring a fraction of sites $\{i_1,i_2,\ldots,i_{K}\}$ in the Rydberg occupation number basis with post-selected outcomes $\mathbf{n} = \{n_{i_1},n_{i_2}, ... ,n_{i_K} \}$. Measurements generically break translation symmetry $T_x$, though we always consider choices of $\mathbf{n}$ that enlarge the unit cell to $p$ sites, preserving a subgroup $(T_x)^p$ on periodic chains; see, e.g., Figs.~\ref{fig:Setup}(b,c).  Bond reflections $R_x$ may either be broken or preserved, as in Figs.~\ref{fig:Setup}(b) and (c), respectively.  The (unnormalized) post-measurement state reads
\begin{equation}
|\psi^{\mathbf{n}}_c\rangle = e^{-\frac{\beta}{2}\hat{\mathcal{H}}_{\mathbf{n}}}|\psi_c\rangle ,~~~~\hat{\mathcal{H}}_{\mathbf{n}} =  \sum\limits_{a=1}^K (-1)^{n_{i_a}}\hat{n}_{i_a}
\label{eq:pmwf1}
\end{equation}
with $\beta$ parametrizing the measurement strength. 
Our experimental protocol and numerical simulations always invoke the projective limit $\beta \rightarrow \infty$, corresponding to $e^{-\frac{\beta}{2}\hat{\mathcal{H}}_{\mathbf{n}}}\propto\prod\limits_{j=1}^K|n_{i_j}\rangle \langle n_{i_j}|$.  We evaluate measurement-induced changes to critical correlations by probing microscopic operators $\hat\sigma_{\ell}^{\bf n}$ and (for protocols that preserve $R_x$) $\hat\varepsilon_{\ell}^{\bf n}$ that respectively map onto the CFT fields $\sigma$ and $\varepsilon$ after measurement; see Appendix~\ref{app:A} for details.  

Complementary analytical insights follow from the weak-measurement limit $0 < \beta \ll 1$.  At $\beta = 0$, expectation values $\bra{\psi_c^{\mathbf{n}}}\cdots \ket{\psi_c^{\mathbf{n}}}$ can be recast as a path integral weighted by a `pure' Euclidean CFT action $S_{\rm CFT}$ defined over space-time coordinates $(x,\tau)$, such that  the imaginary time intervals $\tau > 0$ and $\tau <0$ respectively prepare the quantum critical bra $\bra{\psi_c}$ and ket $\ket{\psi_c}$.  Turning on weak $\beta$ perturbs $S_{\rm CFT}$ with a `defect line' action $S_{\rm meas}$ that acts at all $x$ but only $\tau = 0$ \cite{AltmanMeasurementLL}.  Given measurement data $\mathbf{n}$, the action $S_{\rm meas}$ is constrained by symmetry and explicitly follows from coarse-graining $\beta \hat{\mathcal{H}}_{\mathbf{n}}$ using the field decompositions specified in Eqs.~\eqref{eq:epsilon_map} and \eqref{eq:sigma_map}.  The most general form we will need for either the Ising or TCI cases is
\begin{equation}
    S_{\rm meas} \sim \beta \int_{x} [a_\sigma \sigma(x,\tau = 0) +a_\varepsilon \varepsilon(x,\tau = 0)],
    \label{eq:microscopicdla}
\end{equation}
where we neglected subleading terms, e.g., involving derivatives.  Coefficients $a_\sigma$ and $a_\varepsilon$ are coarse-grained over a length scale large compared to the unit cell size and are therefore position-independent; their magnitudes and relative size do, however, depend on $\mathbf{n}$, and consequently the choice of measured sites and post-selected outcomes controls the defect-line action and consequences thereof.  Measurement patterns that preserve $(T_x)^{p \in {\rm odd}}$ and/or $R_x$, for instance, constrain $a_\sigma = 0$, while otherwise both $a_\sigma$ and $a_\varepsilon$ are generically non-zero. 
Renormalization group and boundary CFT techniques predict modifications of long-distance correlations stemming from weak measurements generating Eq.~\eqref{eq:microscopicdla}. We will see that, in all cases, partial projective measurements implemented by our protocol ($\beta \rightarrow \infty$ at fixed $L$) are consistent with expectations based on these field-theoretic treatments valid at small $\beta$. 

\textbf{\emph{Ising criticality.}}~Suppose now that $\ket{\psi_c}$ represents a quantum critical state tuned to the Ising transition line in Fig.~\ref{fig:Setup}(a).  In the Ising CFT, the $\varepsilon$ term from Eq.~\eqref{eq:microscopicdla} is marginal whereas $\sigma$ is strongly relevant.  Provided $a_\sigma \neq 0$, the latter thus dominates---inducing a flow to a new stable fixed point described by a boundary CFT with fixed boundary conditions at $\tau = 0$ \cite{Cardy1984,Cardy1989}.  Here measurements $(i)$ generate order-parameter condensation, $(ii)$ non-perturbatively modify the $\tau = 0$ $\sigma$ and $\varepsilon$ scaling dimensions to 
$\Delta_\sigma = \Delta_\varepsilon = 2$, and $(iii)$ produce area-law entanglement in the post-measurement wavefunction \cite{usmeasurementaltered, JianMeasurementIsing}.  The marginal $\varepsilon$ term controls the long-distance physics, however, if $a_\sigma$ vanishes due to symmetry; measurements then leave $\Delta_\varepsilon$ unchanged, while both the scaling dimension $\Delta_\sigma$ and effective central charge (quantified by the entanglement entropy) vary continuously with the measurement strength $\beta$ \cite{usmeasurementaltered, JianMeasurementIsing,EhudMeasurementIsing,CABRA_1994,NT2011}.

\begin{figure}
\includegraphics[width=\linewidth]{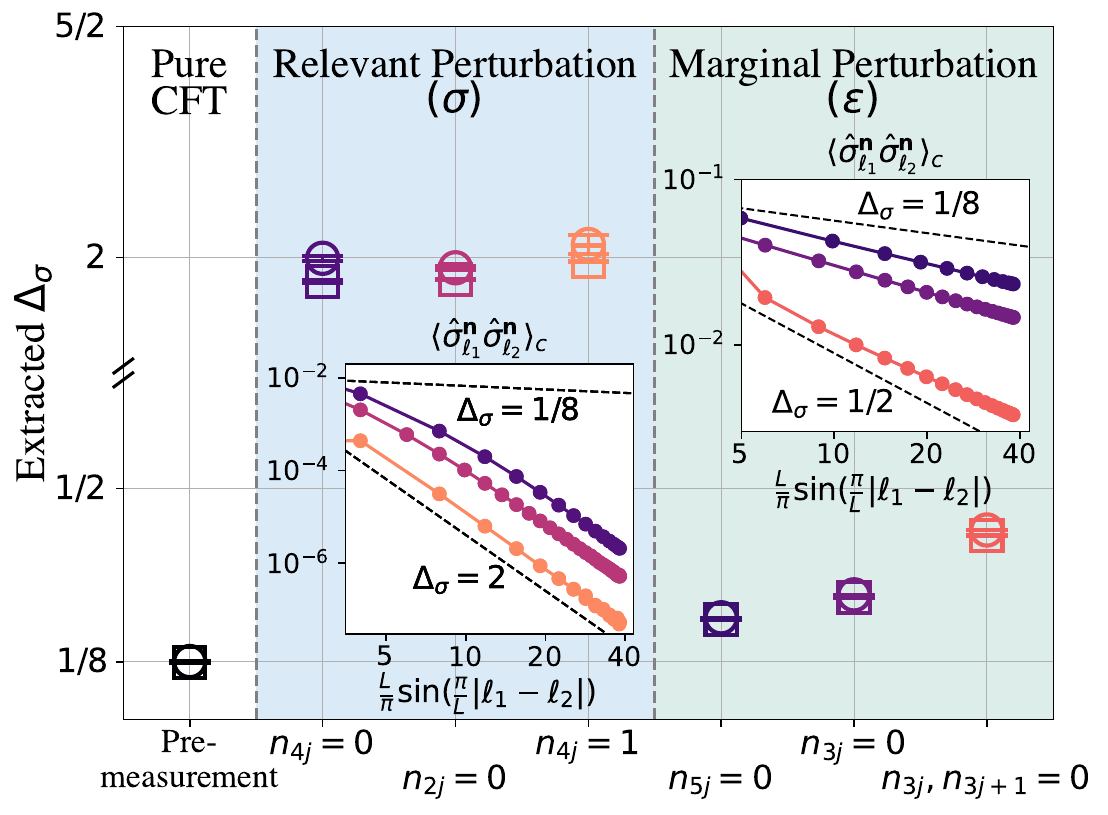}
\caption{\textbf{Altered Ising criticality.} Extracted scaling dimension of $\sigma$ ($\Delta_{\sigma}$) at Ising criticality before (left region) and after (middle and right regions) projective measurements. Horizontal axis indicates various measurement protocols, including the subset of measured sites and their outcomes.  Circles (squares) correspond to chains with periodic (open) boundary conditions. \emph{Insets:} Post-measurement order parameter correlators for periodic chains, which were used to determine the circular markers of the same color in the main panel. The dependence on chord distance arises from the conformal mapping from an infinite line to a circle of circumference $L$ \cite{yellowbook}.
Corresponding open-chain correlators and DMRG implementation details appear in Appendices~\ref{app:C} and~\ref{app:D}.}
\label{fig:Ising_FixedPoints}
\end{figure}

To test these predictions with more practical projective measurements, we employ density matrix renormalization group (DMRG) simulations to extract correlations of a critical Rydberg chain before and after measuring. 
Specifically, we simulate a finite-size chain and perform projective measurements on a subset of sites corresponding to various choices of $\mathbf{n}$ at $V_2 = 0$ 
in periodic chains and with open boundaries.  
Scaling dimensions are extracted from power-law fits of two-point correlators or, with open boundaries, from the decay of one-point $\sigma$ correlators away from edges \cite{Slagle2021}.  
Pre-measurement correlators simply return scaling dimensions in excellent agreement with those of the `pure' Ising CFT---a useful diagnostic for the fidelity of the critical state (see Fig.~\ref{fig:Ising_FixedPoints}, left). The middle (blue) and right (green) regions of Fig.~\ref{fig:Ising_FixedPoints} summarize our post-measurement results for the $\tau = 0$ scaling dimension $\Delta_{\sigma}$ arising in $\mathbf{n}$ sectors specified on the horizontal axis.  (See Appendix~\ref{app:D} for results on $\Delta_\varepsilon$.)   

Middle data correspond to measurements that preserve neither $(T_x)^{p \in {\rm odd}}$ nor $R_x$, thus permitting a nontrivial $\sigma$ perturbation in Eq.~\eqref{eq:microscopicdla}.  For intuition, consider the case where we measure sites $\{2,4,6,\ldots,L\}$ with outcomes $\{0,0,0,\ldots,0\}$, i.e., post-selecting for the atomic ground state on every other site.  Hereafter we adopt shorthand notation $\mathbf{n} = \{n_{2j} = 0\}$ for this protocol (and similarly for other outcomes).  
This measurement sector reduces translation symmetry in a manner that accommodates only one of the two CDW ordered patterns.   
Analogous logic holds for the sectors with $\mathbf{n} = \{n_{4j} = 0\}$ or $\mathbf{n} = \{n_{4j} = 1\}$, albeit with weaker ordering bias because fewer sites are measured.  For all these cases our DMRG simulations recover $\Delta_\sigma \approx 2$,  which agrees with the intuition that these CDW-order-seeding measurements act as a $\sigma$ deformation.

The right section of Fig.~\ref{fig:Ising_FixedPoints} corresponds to measurements that preserve $(T_x)^{p \in {\rm odd}}$ and $R_x$, 
leaving only the $\varepsilon$ perturbation in Eq.~\eqref{eq:microscopicdla}.  Measurements with $\mathbf{n} = \{n_{3j} = 0\}$, for instance, do not discriminate between the two adjacent CDW orders; instead order-parameter correlations are simply suppressed given the ground-state atomic configuration imposed on every third site.  One can control this suppression by choosing $\mathbf{n}$'s with the same symmetry but different densities of measured sites.  Consistent with the intuition above as well as predictions from the weak-measurement limit, our DMRG results reveal $\mathbf{n}$-dependent $\Delta_\sigma$'s exceeding the pre-measurement value of $1/8$.   

\textbf{\emph{Tricritical Ising.}}~Next we examine the critical ground state $\ket{\psi_c}$ at the TCI point.  Here both the $\sigma$ \emph{and} $\varepsilon$ terms from Eq.~\eqref{eq:microscopicdla} constitute strongly relevant perturbations.  When $a_\sigma$ dominates, the system again flows to a stable fixed point with the same characteristics highlighted earlier for the Ising case (including modified $\tau = 0$ dimensions $\Delta_\sigma = \Delta_\varepsilon = 2$).  Measurements yielding dominant $a_\varepsilon$,\footnote{We assume a sign of $a_\varepsilon$ such that the $\varepsilon$ term opposes CDW order, which is always the case for $\mathbf{n}$'s that we have studied.} in contrast, now induce a flow to a different stable fixed point exhibiting free boundary conditions---non-perturbatively modifying the $\tau = 0$ dimensions to $\Delta_\sigma = 3/2, \Delta_\varepsilon = 2$~\cite{FROJDH1991}.  At some critical value of $a_\sigma / a_\varepsilon$, measurements drive a flow to an intervening unstable fixed point   
corresponding to a boundary CFT with  partially polarized boundary conditions and $\Delta_\sigma = \Delta_\varepsilon = 3/5$ \cite{Cardy1989,AffleckTCI_2000}.

\begin{figure}
\includegraphics[width=\linewidth]{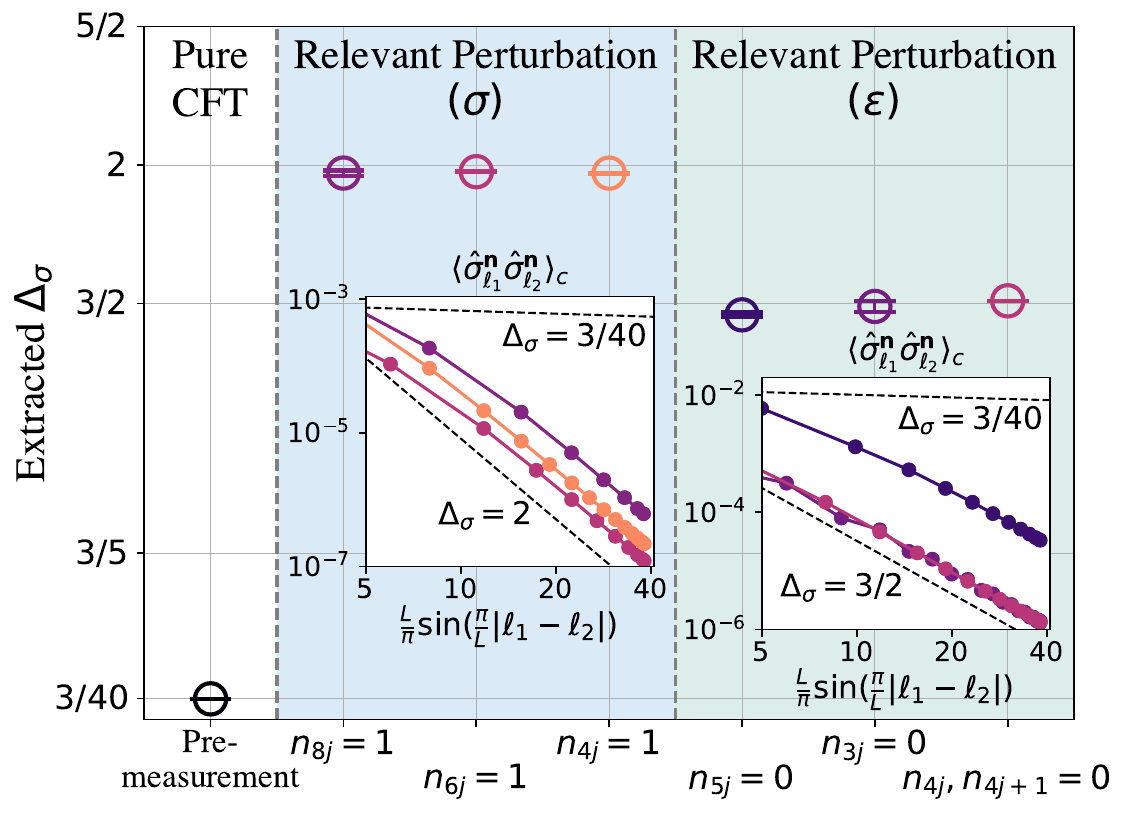}
\caption{\label{fig:TCI_FixedPoints} 
\textbf{Altered tricritical Ising.} Extracted scaling dimension of $\sigma$ ($\Delta_{\sigma}$) at TCI before (left region) and after (middle and right regions) projective measurements (patterns of which are indicated on the horizontal axis) with periodic boundary conditions. \emph{Insets:} Chord distance dependence of the post-measurement order parameter correlators. Additional data for open boundary conditions and the intermediate fixed point can be found in Appendices~\ref{app:C} and~\ref{app:D}, respectively.}
\end{figure} 

Numerical simulations are consistent with these predictions for the TCI chain; see Fig.~\ref{fig:TCI_FixedPoints} for periodic systems (Appendix~\ref{app:D} includes additional details for open boundary conditions). The green region on the right once again corresponds to measurement outcomes for which symmetry dictates $a_\sigma = 0$, in this case yielding $\Delta_\sigma \approx 3/2$ as expected for the TCI point.  The blue region on the left represents outcomes with $n_j = 1$ on every fourth, sixth, or eighth site.  Because of the blockade constraint enforcing $n_{j-1}= n_{j+1} = 0$ on the adjacent sites, such outcomes clearly inject CDW order, acting as a $\sigma$ deformation to indeed yield $\Delta_\sigma \approx 2$ as expected from the $a_\sigma$-driven fixed point.  In principle, one could tune $a_\sigma/a_\varepsilon$ to cross the intermediate unstable fixed point by considering measurement patterns that intermix the two types of outcomes above in varying proportions.  Interestingly, however, Appendix~\ref{app:B} provides numerical evidence that simple outcomes with $n_{2j} = 0$ place the system at or very near the intermediate fixed point. 

\textbf{\emph{Post-selection probabilities.}}~Our approach to observing measurement-altered criticality requires repeatedly obtaining target states $\ket{\psi_c^{\mathbf{n}}}$ by post-selectively measuring a finite fraction $k \equiv K/L$ of sites.  Although the success probability $P_{\mathbf{n}} \sim e^{-k L/\xi_{\mathbf{n}} }$ generically decays exponentially with $L$, we find that $k/\xi_{\mathbf{n}}$ can be quite small for certain outcomes---implying a relatively benign post-selection problem that can be further tamed by reducing the density of measured sites $k$.   

Figure~\ref{fig:PSprobs} presents $P_{\mathbf{n}}$ 
for various $\mathbf{n}$ sectors extracted from DMRG simulations of both critical Ising and TCI wavefunctions.  Panel (a) reveals encouragingly high probabilities for the $\{n_{2j} = 0\}$ outcome even for fairly large systems of order 100 sites, which we attribute to several factors:  
Among protocols that measure every other site, the outcome $\{n_{2j} = 0\}$ maximizes the subspace dimension of the unmeasured sites (any site measured to have $n_j = 1$ pins the adjacent sites to $n_{j-1} = n_{j+1} = 0$ due to Rydberg blockade).  The Fig.~\ref{fig:PSprobs}(a) inset moreover shows that the conditional probability for obtaining $n_{2k} = 0$ given prior measurements of $n_{2k-2} = n_{2k-4} = \cdots = n_0 = 0$ rapidly reaches near unity as $k$ increases---consistent with the observed slow decay with $L$.  Finally, the relatively high probability for $\{n_{2j} = 0\}$ is rather natural given that this outcome accommodates both CDW ($\ldots010101\ldots$) and disordered ($\ldots000000\ldots$) configurations that are associated with the adjacent gapped phases and thus heavily weighted in the critical wavefunction---particularly at the TCI point beyond which the transition becomes first order.  

Probabilities for additional outcomes shown in panel (b) decay much faster with $L$.  Nevertheless, even those outcomes remain feasible for chains with a few tens of sites.  Note that degeneracy factors would enhance their success probability by a few times; e.g., obtaining $\{n_{3j} = 0\}, \{n_{3j+1} = 0\}$, or $\{n_{3j+2} = 0\}$ yield symmetry-equivalent measurement-altered states, at least on a periodic chain.  

{\bf \emph{Discussion.}}~In the pre-measurement quantum critical state $\ket{\psi_c}$, expectation values of generic operators $f(\{\hat{n}_j\})$ diagonal in the number basis can be recast as an unrestricted sum over eigenvalues $f(\{{n}_j\})$ weighted by the corresponding Born probabilities.
To evaluate such quantities in practice, one would repeatedly prepare $\ket{\psi_c}$ and then projectively measure all sites simultaneously to sample the Born probability distribution.

Precisely the same data set encodes measurement-altered correlations. Indeed expectation values in the post-measurement state $\ket{\psi_c^{\mathbf{n}}}$ take an identical form 
but where the sum is restricted over occupation numbers compatible with a target post-selection sector $\mathbf{n}$. 
Additional overhead for probing measurement-altered criticality only comes from an increased number of experimental runs needed to access target sectors with sufficient statistics to back out measurement-altered correlations, whose amplitudes are typically reduced compared to their pre-measurement counterparts. While this increase is generically exponential in system size, we showed that relevant post-selection probabilities are enhanced in part by the Rydberg blockade.
Another virtue of the Rydberg setup is that measurements in a \emph{single basis} suffice to not only probe different CFT fields [recall Eqs.~\eqref{eq:epsilon_map} and \eqref{eq:sigma_map}], but also to access distinct measurement-induced modifications by simply tailoring the post-selected outcomes $\mathbf{n}$.  This feature reflects the fact that the CDW ordered states break translation symmetry, rather than a purely on-site symmetry.  The transverse-field Ising model with local spin-flip symmetry, for instance, would require measurements in two different bases \cite{usmeasurementaltered,EhudMeasurementIsing,JianMeasurementIsing}.  

Looking ahead, it will be important to characterize decoherence and finite-temperature effects occurring in the experimental setup, as well as possible error correction and decoding protocols to further facilitate the demonstration of measurement-altered criticality. The Rydberg setup may further allow realizing predictions from Ref.~\onlinecite{Ludwig2024HighlyComplex} on the same CFTs studied here, regarding non-linear quantities in the density matrix; devising practical algorithms to that end poses another interesting future direction. \\

{\bf \emph{Acknowledgments.}}~It is a pleasure to acknowledge illuminating conversations with Natalia Chepiga, Paul Fendley, Samuel Garratt, Johannes Hauschild, Nandagopal Manoj, Xiangkai Sun, and Zack Weinstein. This work was primarily supported by the U.S.~Department of Energy, Office of Science, National Quantum Information Science Research Centers, Quantum Science Center and Quantum Systems Accelerator. Tensor network calculations
were performed using the TeNPy Library~\cite{tenpy}.
 Additional support was provided by the Institute for Quantum Information and Matter, an NSF Physics Frontiers Center (PHY-2317110); and the Walter Burke Institute for Theoretical Physics at Caltech.
 

\begin{thebibliography}{40}%
\makeatletter
\providecommand \@ifxundefined [1]{%
 \@ifx{#1\undefined}
}%
\providecommand \@ifnum [1]{%
 \ifnum #1\expandafter \@firstoftwo
 \else \expandafter \@secondoftwo
 \fi
}%
\providecommand \@ifx [1]{%
 \ifx #1\expandafter \@firstoftwo
 \else \expandafter \@secondoftwo
 \fi
}%
\providecommand \natexlab [1]{#1}%
\providecommand \enquote  [1]{``#1''}%
\providecommand \bibnamefont  [1]{#1}%
\providecommand \bibfnamefont [1]{#1}%
\providecommand \citenamefont [1]{#1}%
\providecommand \href@noop [0]{\@secondoftwo}%
\providecommand \href [0]{\begingroup \@sanitize@url \@href}%
\providecommand \@href[1]{\@@startlink{#1}\@@href}%
\providecommand \@@href[1]{\endgroup#1\@@endlink}%
\providecommand \@sanitize@url [0]{\catcode `\\12\catcode `\$12\catcode `\&12\catcode `\#12\catcode `\^12\catcode `\_12\catcode `\%12\relax}%
\providecommand \@@startlink[1]{}%
\providecommand \@@endlink[0]{}%
\providecommand \url  [0]{\begingroup\@sanitize@url \@url }%
\providecommand \@url [1]{\endgroup\@href {#1}{\urlprefix }}%
\providecommand \urlprefix  [0]{URL }%
\providecommand \Eprint [0]{\href }%
\providecommand \doibase [0]{https://doi.org/}%
\providecommand \selectlanguage [0]{\@gobble}%
\providecommand \bibinfo  [0]{\@secondoftwo}%
\providecommand \bibfield  [0]{\@secondoftwo}%
\providecommand \translation [1]{[#1]}%
\providecommand \BibitemOpen [0]{}%
\providecommand \bibitemStop [0]{}%
\providecommand \bibitemNoStop [0]{.\EOS\space}%
\providecommand \EOS [0]{\spacefactor3000\relax}%
\providecommand \BibitemShut  [1]{\csname bibitem#1\endcsname}%
\let\auto@bib@innerbib\@empty
\bibitem [{\citenamefont {Garratt}\ \emph {et~al.}(2023)\citenamefont {Garratt}, \citenamefont {Weinstein},\ and\ \citenamefont {Altman}}]{AltmanMeasurementLL}%
  \BibitemOpen
  \bibfield  {author} {\bibinfo {author} {\bibfnamefont {S.~J.}\ \bibnamefont {Garratt}}, \bibinfo {author} {\bibfnamefont {Z.}~\bibnamefont {Weinstein}},\ and\ \bibinfo {author} {\bibfnamefont {E.}~\bibnamefont {Altman}},\ }\href {https://doi.org/10.1103/PhysRevX.13.021026} {\bibfield  {journal} {\bibinfo  {journal} {Phys. Rev. X}\ }\textbf {\bibinfo {volume} {13}},\ \bibinfo {pages} {021026} (\bibinfo {year} {2023})}\BibitemShut {NoStop}%
\bibitem [{\citenamefont {Sun}\ \emph {et~al.}(2023)\citenamefont {Sun}, \citenamefont {Yao},\ and\ \citenamefont {Jian}}]{sun2023}%
  \BibitemOpen
  \bibfield  {author} {\bibinfo {author} {\bibfnamefont {X.}~\bibnamefont {Sun}}, \bibinfo {author} {\bibfnamefont {H.}~\bibnamefont {Yao}},\ and\ \bibinfo {author} {\bibfnamefont {S.-K.}\ \bibnamefont {Jian}},\ }\href@noop {} {\bibinfo {title} {New critical states induced by measurement}} (\bibinfo {year} {2023}),\ \Eprint {https://arxiv.org/abs/2301.11337} {arXiv:2301.11337 [quant-ph]} \BibitemShut {NoStop}%
\bibitem [{\citenamefont {Ashida}\ \emph {et~al.}(2023)\citenamefont {Ashida}, \citenamefont {Furukawa},\ and\ \citenamefont {Oshikawa}}]{Ashida23}%
  \BibitemOpen
  \bibfield  {author} {\bibinfo {author} {\bibfnamefont {Y.}~\bibnamefont {Ashida}}, \bibinfo {author} {\bibfnamefont {S.}~\bibnamefont {Furukawa}},\ and\ \bibinfo {author} {\bibfnamefont {M.}~\bibnamefont {Oshikawa}},\ }\href@noop {} {\bibinfo {title} {System-environment entanglement phase transitions}} (\bibinfo {year} {2023}),\ \Eprint {https://arxiv.org/abs/2311.16343} {arXiv:2311.16343 [cond-mat.stat-mech]} \BibitemShut {NoStop}%
\bibitem [{\citenamefont {Weinstein}\ \emph {et~al.}(2023)\citenamefont {Weinstein}, \citenamefont {Sajith}, \citenamefont {Altman},\ and\ \citenamefont {Garratt}}]{EhudMeasurementIsing}%
  \BibitemOpen
  \bibfield  {author} {\bibinfo {author} {\bibfnamefont {Z.}~\bibnamefont {Weinstein}}, \bibinfo {author} {\bibfnamefont {R.}~\bibnamefont {Sajith}}, \bibinfo {author} {\bibfnamefont {E.}~\bibnamefont {Altman}},\ and\ \bibinfo {author} {\bibfnamefont {S.~J.}\ \bibnamefont {Garratt}},\ }\href {https://doi.org/10.1103%2Fphysrevb.107.245132} {\bibfield  {journal} {\bibinfo  {journal} {Phys. Rev. B}\ }\textbf {\bibinfo {volume} {107}},\ \bibinfo {pages} {245132} (\bibinfo {year} {2023})}\BibitemShut {NoStop}%
\bibitem [{\citenamefont {Yang}\ \emph {et~al.}(2023)\citenamefont {Yang}, \citenamefont {Mao},\ and\ \citenamefont {Jian}}]{JianMeasurementIsing}%
  \BibitemOpen
  \bibfield  {author} {\bibinfo {author} {\bibfnamefont {Z.}~\bibnamefont {Yang}}, \bibinfo {author} {\bibfnamefont {D.}~\bibnamefont {Mao}},\ and\ \bibinfo {author} {\bibfnamefont {C.-M.}\ \bibnamefont {Jian}},\ }\href {https://doi.org/10.1103/PhysRevB.108.165120} {\bibfield  {journal} {\bibinfo  {journal} {Phys. Rev. B}\ }\textbf {\bibinfo {volume} {108}},\ \bibinfo {pages} {165120} (\bibinfo {year} {2023})}\BibitemShut {NoStop}%
\bibitem [{\citenamefont {Murciano}\ \emph {et~al.}(2023)\citenamefont {Murciano}, \citenamefont {Sala}, \citenamefont {Liu}, \citenamefont {Mong},\ and\ \citenamefont {Alicea}}]{usmeasurementaltered}%
  \BibitemOpen
  \bibfield  {author} {\bibinfo {author} {\bibfnamefont {S.}~\bibnamefont {Murciano}}, \bibinfo {author} {\bibfnamefont {P.}~\bibnamefont {Sala}}, \bibinfo {author} {\bibfnamefont {Y.}~\bibnamefont {Liu}}, \bibinfo {author} {\bibfnamefont {R.~S.~K.}\ \bibnamefont {Mong}},\ and\ \bibinfo {author} {\bibfnamefont {J.}~\bibnamefont {Alicea}},\ }\href {https://doi.org/10.1103/PhysRevX.13.041042} {\bibfield  {journal} {\bibinfo  {journal} {Phys. Rev. X}\ }\textbf {\bibinfo {volume} {13}},\ \bibinfo {pages} {041042} (\bibinfo {year} {2023})}\BibitemShut {NoStop}%
\bibitem [{\citenamefont {Paviglianiti}\ \emph {et~al.}(2023)\citenamefont {Paviglianiti}, \citenamefont {Turkeshi}, \citenamefont {Schir\`o},\ and\ \citenamefont {Silva}}]{Paviglianiti2023}%
  \BibitemOpen
  \bibfield  {author} {\bibinfo {author} {\bibfnamefont {A.}~\bibnamefont {Paviglianiti}}, \bibinfo {author} {\bibfnamefont {X.}~\bibnamefont {Turkeshi}}, \bibinfo {author} {\bibfnamefont {M.}~\bibnamefont {Schir\`o}},\ and\ \bibinfo {author} {\bibfnamefont {A.}~\bibnamefont {Silva}},\ }\href@noop {} {\bibinfo {title} {{Enhanced Entanglement in the Measurement-Altered Quantum Ising Chain}}} (\bibinfo {year} {2023}),\ \Eprint {https://arxiv.org/abs/2310.02686} {arXiv:2310.02686 [quant-ph]} \BibitemShut {NoStop}%
\bibitem [{\citenamefont {Patil}\ and\ \citenamefont {Ludwig}(2024)}]{Ludwig2024HighlyComplex}%
  \BibitemOpen
  \bibfield  {author} {\bibinfo {author} {\bibfnamefont {R.~A.}\ \bibnamefont {Patil}}\ and\ \bibinfo {author} {\bibfnamefont {A.~W.~W.}\ \bibnamefont {Ludwig}},\ }\href {https://arxiv.org/abs/2409.02107} {\bibinfo {title} {Highly complex novel critical behavior from the intrinsic randomness of quantum mechanical measurements on critical ground states -- a controlled renormalization group analysis}} (\bibinfo {year} {2024}),\ \Eprint {https://arxiv.org/abs/2409.02107} {arXiv:2409.02107 [cond-mat.stat-mech]} \BibitemShut {NoStop}%
\bibitem [{\citenamefont {Lee}\ \emph {et~al.}(2023)\citenamefont {Lee}, \citenamefont {Jian},\ and\ \citenamefont {Xu}}]{Lee2023}%
  \BibitemOpen
  \bibfield  {author} {\bibinfo {author} {\bibfnamefont {J.~Y.}\ \bibnamefont {Lee}}, \bibinfo {author} {\bibfnamefont {C.-M.}\ \bibnamefont {Jian}},\ and\ \bibinfo {author} {\bibfnamefont {C.}~\bibnamefont {Xu}},\ }\href {https://doi.org/10.1103/PRXQuantum.4.030317} {\bibfield  {journal} {\bibinfo  {journal} {PRX Quantum}\ }\textbf {\bibinfo {volume} {4}},\ \bibinfo {pages} {030317} (\bibinfo {year} {2023})}\BibitemShut {NoStop}%
\bibitem [{\citenamefont {Tang}\ and\ \citenamefont {Wen}(2024)}]{tang2024}%
  \BibitemOpen
  \bibfield  {author} {\bibinfo {author} {\bibfnamefont {Q.}~\bibnamefont {Tang}}\ and\ \bibinfo {author} {\bibfnamefont {X.}~\bibnamefont {Wen}},\ }\href {https://arxiv.org/abs/2411.13705} {\bibinfo {title} {A critical state under weak measurement is not critical}} (\bibinfo {year} {2024}),\ \Eprint {https://arxiv.org/abs/2411.13705} {arXiv:2411.13705 [cond-mat.stat-mech]} \BibitemShut {NoStop}%
\bibitem [{\citenamefont {Liu}\ \emph {et~al.}(2024)\citenamefont {Liu}, \citenamefont {Murciano}, \citenamefont {Mross},\ and\ \citenamefont {Alicea}}]{Yue_gapless2024}%
  \BibitemOpen
  \bibfield  {author} {\bibinfo {author} {\bibfnamefont {Y.}~\bibnamefont {Liu}}, \bibinfo {author} {\bibfnamefont {S.}~\bibnamefont {Murciano}}, \bibinfo {author} {\bibfnamefont {D.~F.}\ \bibnamefont {Mross}},\ and\ \bibinfo {author} {\bibfnamefont {J.}~\bibnamefont {Alicea}},\ }\href {https://arxiv.org/abs/2412.07830} {\bibinfo {title} {Boundary transitions from a single round of measurements on gapless quantum states}} (\bibinfo {year} {2024}),\ \Eprint {https://arxiv.org/abs/2412.07830} {arXiv:2412.07830 [quant-ph]} \BibitemShut {NoStop}%
\bibitem [{\citenamefont {Zou}\ \emph {et~al.}(2023)\citenamefont {Zou}, \citenamefont {Sang},\ and\ \citenamefont {Hsieh}}]{Zou2023}%
  \BibitemOpen
  \bibfield  {author} {\bibinfo {author} {\bibfnamefont {Y.}~\bibnamefont {Zou}}, \bibinfo {author} {\bibfnamefont {S.}~\bibnamefont {Sang}},\ and\ \bibinfo {author} {\bibfnamefont {T.~H.}\ \bibnamefont {Hsieh}},\ }\href {https://doi.org/10.1103/PhysRevLett.130.250403} {\bibfield  {journal} {\bibinfo  {journal} {Phys. Rev. Lett.}\ }\textbf {\bibinfo {volume} {130}},\ \bibinfo {pages} {250403} (\bibinfo {year} {2023})}\BibitemShut {NoStop}%
\bibitem [{\citenamefont {Sala}\ \emph {et~al.}(2024)\citenamefont {Sala}, \citenamefont {Murciano}, \citenamefont {Liu},\ and\ \citenamefont {Alicea}}]{Qteleport}%
  \BibitemOpen
  \bibfield  {author} {\bibinfo {author} {\bibfnamefont {P.}~\bibnamefont {Sala}}, \bibinfo {author} {\bibfnamefont {S.}~\bibnamefont {Murciano}}, \bibinfo {author} {\bibfnamefont {Y.}~\bibnamefont {Liu}},\ and\ \bibinfo {author} {\bibfnamefont {J.}~\bibnamefont {Alicea}},\ }\href {https://doi.org/10.1103/PRXQuantum.5.030307} {\bibfield  {journal} {\bibinfo  {journal} {PRX Quantum}\ }\textbf {\bibinfo {volume} {5}},\ \bibinfo {pages} {030307} (\bibinfo {year} {2024})}\BibitemShut {NoStop}%
\bibitem [{\citenamefont {Eckstein}\ \emph {et~al.}(2024)\citenamefont {Eckstein}, \citenamefont {Han}, \citenamefont {Trebst},\ and\ \citenamefont {Zhu}}]{GuoYiTeleportation}%
  \BibitemOpen
  \bibfield  {author} {\bibinfo {author} {\bibfnamefont {F.}~\bibnamefont {Eckstein}}, \bibinfo {author} {\bibfnamefont {B.}~\bibnamefont {Han}}, \bibinfo {author} {\bibfnamefont {S.}~\bibnamefont {Trebst}},\ and\ \bibinfo {author} {\bibfnamefont {G.-Y.}\ \bibnamefont {Zhu}},\ }\href {https://doi.org/10.1103/PRXQuantum.5.040313} {\bibfield  {journal} {\bibinfo  {journal} {PRX Quantum}\ }\textbf {\bibinfo {volume} {5}},\ \bibinfo {pages} {040313} (\bibinfo {year} {2024})}\BibitemShut {NoStop}%
\bibitem [{\citenamefont {Hoshino}\ \emph {et~al.}(2025)\citenamefont {Hoshino}, \citenamefont {Oshikawa},\ and\ \citenamefont {Ashida}}]{Hoshino2024}%
  \BibitemOpen
  \bibfield  {author} {\bibinfo {author} {\bibfnamefont {M.}~\bibnamefont {Hoshino}}, \bibinfo {author} {\bibfnamefont {M.}~\bibnamefont {Oshikawa}},\ and\ \bibinfo {author} {\bibfnamefont {Y.}~\bibnamefont {Ashida}},\ }\href {https://doi.org/10.1103/PhysRevB.111.155143} {\bibfield  {journal} {\bibinfo  {journal} {Phys. Rev. B}\ }\textbf {\bibinfo {volume} {111}},\ \bibinfo {pages} {155143} (\bibinfo {year} {2025})},\ \Eprint {https://arxiv.org/abs/2406.12377} {arXiv:2406.12377 [quant-ph]} \BibitemShut {NoStop}%
\bibitem [{\citenamefont {Nahum}\ and\ \citenamefont {Jacobsen}(2025)}]{Nahum2025}%
  \BibitemOpen
  \bibfield  {author} {\bibinfo {author} {\bibfnamefont {A.}~\bibnamefont {Nahum}}\ and\ \bibinfo {author} {\bibfnamefont {J.~L.}\ \bibnamefont {Jacobsen}},\ }\href@noop {} {\bibinfo {title} {{Bayesian critical points in classical lattice models}}} (\bibinfo {year} {2025}),\ \Eprint {https://arxiv.org/abs/2504.01264} {arXiv:2504.01264 [cond-mat.stat-mech]} \BibitemShut {NoStop}%
\bibitem [{\citenamefont {P\"utz}\ \emph {et~al.}(2025)\citenamefont {P\"utz}, \citenamefont {Garratt}, \citenamefont {Nishimori}, \citenamefont {Trebst},\ and\ \citenamefont {Zhu}}]{Putz2025}%
  \BibitemOpen
  \bibfield  {author} {\bibinfo {author} {\bibfnamefont {M.}~\bibnamefont {P\"utz}}, \bibinfo {author} {\bibfnamefont {S.~J.}\ \bibnamefont {Garratt}}, \bibinfo {author} {\bibfnamefont {H.}~\bibnamefont {Nishimori}}, \bibinfo {author} {\bibfnamefont {S.}~\bibnamefont {Trebst}},\ and\ \bibinfo {author} {\bibfnamefont {G.-Y.}\ \bibnamefont {Zhu}},\ }\href@noop {} {\bibinfo {title} {{Learning transitions in classical Ising models and deformed toric codes}}} (\bibinfo {year} {2025}),\ \Eprint {https://arxiv.org/abs/2504.12385} {arXiv:2504.12385 [cond-mat.stat-mech]} \BibitemShut {NoStop}%
\bibitem [{\citenamefont {{Garratt}}\ and\ \citenamefont {{Altman}}(2023)}]{Garratt23}%
  \BibitemOpen
  \bibfield  {author} {\bibinfo {author} {\bibfnamefont {S.~J.}\ \bibnamefont {{Garratt}}}\ and\ \bibinfo {author} {\bibfnamefont {E.}~\bibnamefont {{Altman}}},\ }\href {https://doi.org/10.48550/arXiv.2305.20092} {\bibfield  {journal} {\bibinfo  {journal} {arXiv e-prints}\ ,\ \bibinfo {eid} {arXiv:2305.20092}} (\bibinfo {year} {2023})},\ \Eprint {https://arxiv.org/abs/2305.20092} {arXiv:2305.20092 [quant-ph]} \BibitemShut {NoStop}%
\bibitem [{\citenamefont {Fendley}\ \emph {et~al.}(2004)\citenamefont {Fendley}, \citenamefont {Sengupta},\ and\ \citenamefont {Sachdev}}]{FSS}%
  \BibitemOpen
  \bibfield  {author} {\bibinfo {author} {\bibfnamefont {P.}~\bibnamefont {Fendley}}, \bibinfo {author} {\bibfnamefont {K.}~\bibnamefont {Sengupta}},\ and\ \bibinfo {author} {\bibfnamefont {S.}~\bibnamefont {Sachdev}},\ }\href {https://doi.org/10.1103/PhysRevB.69.075106} {\bibfield  {journal} {\bibinfo  {journal} {Phys. Rev. B}\ }\textbf {\bibinfo {volume} {69}},\ \bibinfo {pages} {075106} (\bibinfo {year} {2004})}\BibitemShut {NoStop}%
\bibitem [{\citenamefont {Bernien}\ \emph {et~al.}(2017)\citenamefont {Bernien}, \citenamefont {Schwartz}, \citenamefont {Keesling}, \citenamefont {Levine}, \citenamefont {Omran}, \citenamefont {Pichler}, \citenamefont {Choi}, \citenamefont {Zibrov}, \citenamefont {Endres}, \citenamefont {Greiner}, \citenamefont {Vuletić},\ and\ \citenamefont {Lukin}}]{Bernien_2017}%
  \BibitemOpen
  \bibfield  {author} {\bibinfo {author} {\bibfnamefont {H.}~\bibnamefont {Bernien}}, \bibinfo {author} {\bibfnamefont {S.}~\bibnamefont {Schwartz}}, \bibinfo {author} {\bibfnamefont {A.}~\bibnamefont {Keesling}}, \bibinfo {author} {\bibfnamefont {H.}~\bibnamefont {Levine}}, \bibinfo {author} {\bibfnamefont {A.}~\bibnamefont {Omran}}, \bibinfo {author} {\bibfnamefont {H.}~\bibnamefont {Pichler}}, \bibinfo {author} {\bibfnamefont {S.}~\bibnamefont {Choi}}, \bibinfo {author} {\bibfnamefont {A.~S.}\ \bibnamefont {Zibrov}}, \bibinfo {author} {\bibfnamefont {M.}~\bibnamefont {Endres}}, \bibinfo {author} {\bibfnamefont {M.}~\bibnamefont {Greiner}}, \bibinfo {author} {\bibfnamefont {V.}~\bibnamefont {Vuletić}},\ and\ \bibinfo {author} {\bibfnamefont {M.~D.}\ \bibnamefont {Lukin}},\ }\href {https://doi.org/10.1038/nature24622} {\bibfield  {journal} {\bibinfo  {journal} {Nature}\ }\textbf {\bibinfo {volume} {551}},\ \bibinfo {pages} {579–584} (\bibinfo {year} {2017})}\BibitemShut {NoStop}%
\bibitem [{\citenamefont {Keesling}\ \emph {et~al.}(2019)\citenamefont {Keesling}, \citenamefont {Omran}, \citenamefont {Levine}, \citenamefont {Bernien}, \citenamefont {Pichler}, \citenamefont {Choi}, \citenamefont {Samajdar}, \citenamefont {Schwartz}, \citenamefont {Silvi}, \citenamefont {Sachdev}, \citenamefont {Zoller}, \citenamefont {Endres}, \citenamefont {Greiner}, \citenamefont {Vuletić},\ and\ \citenamefont {Lukin}}]{Keesling_2019}%
  \BibitemOpen
  \bibfield  {author} {\bibinfo {author} {\bibfnamefont {A.}~\bibnamefont {Keesling}}, \bibinfo {author} {\bibfnamefont {A.}~\bibnamefont {Omran}}, \bibinfo {author} {\bibfnamefont {H.}~\bibnamefont {Levine}}, \bibinfo {author} {\bibfnamefont {H.}~\bibnamefont {Bernien}}, \bibinfo {author} {\bibfnamefont {H.}~\bibnamefont {Pichler}}, \bibinfo {author} {\bibfnamefont {S.}~\bibnamefont {Choi}}, \bibinfo {author} {\bibfnamefont {R.}~\bibnamefont {Samajdar}}, \bibinfo {author} {\bibfnamefont {S.}~\bibnamefont {Schwartz}}, \bibinfo {author} {\bibfnamefont {P.}~\bibnamefont {Silvi}}, \bibinfo {author} {\bibfnamefont {S.}~\bibnamefont {Sachdev}}, \bibinfo {author} {\bibfnamefont {P.}~\bibnamefont {Zoller}}, \bibinfo {author} {\bibfnamefont {M.}~\bibnamefont {Endres}}, \bibinfo {author} {\bibfnamefont {M.}~\bibnamefont {Greiner}}, \bibinfo {author} {\bibfnamefont {V.}~\bibnamefont {Vuletić}},\ and\ \bibinfo {author} {\bibfnamefont {M.~D.}\ \bibnamefont {Lukin}},\ }\href {https://doi.org/10.1038/s41586-019-1070-1}
  {\bibfield  {journal} {\bibinfo  {journal} {Nature}\ }\textbf {\bibinfo {volume} {568}},\ \bibinfo {pages} {207–211} (\bibinfo {year} {2019})}\BibitemShut {NoStop}%
\bibitem [{\citenamefont {Slagle}\ \emph {et~al.}(2021)\citenamefont {Slagle}, \citenamefont {Aasen}, \citenamefont {Pichler}, \citenamefont {Mong}, \citenamefont {Fendley}, \citenamefont {Chen}, \citenamefont {Endres},\ and\ \citenamefont {Alicea}}]{Slagle2021}%
  \BibitemOpen
  \bibfield  {author} {\bibinfo {author} {\bibfnamefont {K.}~\bibnamefont {Slagle}}, \bibinfo {author} {\bibfnamefont {D.}~\bibnamefont {Aasen}}, \bibinfo {author} {\bibfnamefont {H.}~\bibnamefont {Pichler}}, \bibinfo {author} {\bibfnamefont {R.~S.~K.}\ \bibnamefont {Mong}}, \bibinfo {author} {\bibfnamefont {P.}~\bibnamefont {Fendley}}, \bibinfo {author} {\bibfnamefont {X.}~\bibnamefont {Chen}}, \bibinfo {author} {\bibfnamefont {M.}~\bibnamefont {Endres}},\ and\ \bibinfo {author} {\bibfnamefont {J.}~\bibnamefont {Alicea}},\ }\href {https://doi.org/10.1103/PhysRevB.104.235109} {\bibfield  {journal} {\bibinfo  {journal} {Phys. Rev. B}\ }\textbf {\bibinfo {volume} {104}},\ \bibinfo {pages} {235109} (\bibinfo {year} {2021})}\BibitemShut {NoStop}%
\bibitem [{\citenamefont {Scholl}\ \emph {et~al.}(2023)\citenamefont {Scholl}, \citenamefont {Shaw}, \citenamefont {Tsai}, \citenamefont {Finkelstein}, \citenamefont {Choi},\ and\ \citenamefont {Endres}}]{Scholl_2023}%
  \BibitemOpen
  \bibfield  {author} {\bibinfo {author} {\bibfnamefont {P.}~\bibnamefont {Scholl}}, \bibinfo {author} {\bibfnamefont {A.~L.}\ \bibnamefont {Shaw}}, \bibinfo {author} {\bibfnamefont {R.~B.-S.}\ \bibnamefont {Tsai}}, \bibinfo {author} {\bibfnamefont {R.}~\bibnamefont {Finkelstein}}, \bibinfo {author} {\bibfnamefont {J.}~\bibnamefont {Choi}},\ and\ \bibinfo {author} {\bibfnamefont {M.}~\bibnamefont {Endres}},\ }\href {https://doi.org/10.1038/s41586-023-06516-4} {\bibfield  {journal} {\bibinfo  {journal} {Nature}\ }\textbf {\bibinfo {volume} {622}},\ \bibinfo {pages} {273–278} (\bibinfo {year} {2023})}\BibitemShut {NoStop}%
\bibitem [{\citenamefont {Fang}\ \emph {et~al.}(2024)\citenamefont {Fang}, \citenamefont {Wang}, \citenamefont {Liu}, \citenamefont {Wang}, \citenamefont {Cimmino}, \citenamefont {Wei}, \citenamefont {Bintz}, \citenamefont {Parr}, \citenamefont {Kemp}, \citenamefont {Ni},\ and\ \citenamefont {Yao}}]{Fang2024expcrit}%
  \BibitemOpen
  \bibfield  {author} {\bibinfo {author} {\bibfnamefont {F.}~\bibnamefont {Fang}}, \bibinfo {author} {\bibfnamefont {K.}~\bibnamefont {Wang}}, \bibinfo {author} {\bibfnamefont {V.~S.}\ \bibnamefont {Liu}}, \bibinfo {author} {\bibfnamefont {Y.}~\bibnamefont {Wang}}, \bibinfo {author} {\bibfnamefont {R.}~\bibnamefont {Cimmino}}, \bibinfo {author} {\bibfnamefont {J.}~\bibnamefont {Wei}}, \bibinfo {author} {\bibfnamefont {M.}~\bibnamefont {Bintz}}, \bibinfo {author} {\bibfnamefont {A.}~\bibnamefont {Parr}}, \bibinfo {author} {\bibfnamefont {J.}~\bibnamefont {Kemp}}, \bibinfo {author} {\bibfnamefont {K.-K.}\ \bibnamefont {Ni}},\ and\ \bibinfo {author} {\bibfnamefont {N.~Y.}\ \bibnamefont {Yao}},\ }\href {https://arxiv.org/abs/2402.15376} {\bibinfo {title} {Probing critical phenomena in open quantum systems using atom arrays}} (\bibinfo {year} {2024}),\ \Eprint {https://arxiv.org/abs/2402.15376} {arXiv:2402.15376 [quant-ph]} \BibitemShut {NoStop}%
\bibitem [{\citenamefont {Whitsitt}\ \emph {et~al.}(2018)\citenamefont {Whitsitt}, \citenamefont {Samajdar},\ and\ \citenamefont {Sachdev}}]{Whitsitt2018}%
  \BibitemOpen
  \bibfield  {author} {\bibinfo {author} {\bibfnamefont {S.}~\bibnamefont {Whitsitt}}, \bibinfo {author} {\bibfnamefont {R.}~\bibnamefont {Samajdar}},\ and\ \bibinfo {author} {\bibfnamefont {S.}~\bibnamefont {Sachdev}},\ }\bibfield  {journal} {\bibinfo  {journal} {Physical Review. B}\ }\textbf {\bibinfo {volume} {98}},\ \href {https://doi.org/10.1103/physrevb.98.205118} {10.1103/physrevb.98.205118} (\bibinfo {year} {2018})\BibitemShut {NoStop}%
\bibitem [{\citenamefont {Chepiga}\ and\ \citenamefont {Mila}(2019)}]{Chepiga2019}%
  \BibitemOpen
  \bibfield  {author} {\bibinfo {author} {\bibfnamefont {N.}~\bibnamefont {Chepiga}}\ and\ \bibinfo {author} {\bibfnamefont {F.}~\bibnamefont {Mila}},\ }\href {https://doi.org/10.1103/PhysRevLett.122.017205} {\bibfield  {journal} {\bibinfo  {journal} {Phys. Rev. Lett.}\ }\textbf {\bibinfo {volume} {122}},\ \bibinfo {pages} {017205} (\bibinfo {year} {2019})}\BibitemShut {NoStop}%
\bibitem [{\citenamefont {Chepiga}\ and\ \citenamefont {Mila}(2021)}]{Chepiga2021}%
  \BibitemOpen
  \bibfield  {author} {\bibinfo {author} {\bibfnamefont {N.}~\bibnamefont {Chepiga}}\ and\ \bibinfo {author} {\bibfnamefont {F.}~\bibnamefont {Mila}},\ }\bibfield  {journal} {\bibinfo  {journal} {Nature Communications}\ }\textbf {\bibinfo {volume} {12}},\ \href {https://doi.org/10.1038/s41467-020-20641-y} {10.1038/s41467-020-20641-y} (\bibinfo {year} {2021})\BibitemShut {NoStop}%
\bibitem [{\citenamefont {Samajdar}\ \emph {et~al.}(2018)\citenamefont {Samajdar}, \citenamefont {Choi}, \citenamefont {Pichler}, \citenamefont {Lukin},\ and\ \citenamefont {Sachdev}}]{Z3Rydberg_Sachdev}%
  \BibitemOpen
  \bibfield  {author} {\bibinfo {author} {\bibfnamefont {R.}~\bibnamefont {Samajdar}}, \bibinfo {author} {\bibfnamefont {S.}~\bibnamefont {Choi}}, \bibinfo {author} {\bibfnamefont {H.}~\bibnamefont {Pichler}}, \bibinfo {author} {\bibfnamefont {M.~D.}\ \bibnamefont {Lukin}},\ and\ \bibinfo {author} {\bibfnamefont {S.}~\bibnamefont {Sachdev}},\ }\href {https://doi.org/10.1103/PhysRevA.98.023614} {\bibfield  {journal} {\bibinfo  {journal} {Phys. Rev. A}\ }\textbf {\bibinfo {volume} {98}},\ \bibinfo {pages} {023614} (\bibinfo {year} {2018})}\BibitemShut {NoStop}%
\bibitem [{\citenamefont {Chepiga}(2024)}]{Chepiga_2024}%
  \BibitemOpen
  \bibfield  {author} {\bibinfo {author} {\bibfnamefont {N.}~\bibnamefont {Chepiga}},\ }\bibfield  {journal} {\bibinfo  {journal} {Physical Review Letters}\ }\textbf {\bibinfo {volume} {132}},\ \href {https://doi.org/10.1103/physrevlett.132.076505} {10.1103/physrevlett.132.076505} (\bibinfo {year} {2024})\BibitemShut {NoStop}%
\bibitem [{\citenamefont {Zhang}\ \emph {et~al.}(2025)\citenamefont {Zhang}, \citenamefont {Cant{\'u}}, \citenamefont {Liu}, \citenamefont {Bylinskii}, \citenamefont {Braverman}, \citenamefont {Huber}, \citenamefont {Amato-Grill}, \citenamefont {Lukin}, \citenamefont {Gemelke}, \citenamefont {Keesling} \emph {et~al.}}]{zhang2025probing}%
  \BibitemOpen
  \bibfield  {author} {\bibinfo {author} {\bibfnamefont {J.}~\bibnamefont {Zhang}}, \bibinfo {author} {\bibfnamefont {S.~H.}\ \bibnamefont {Cant{\'u}}}, \bibinfo {author} {\bibfnamefont {F.}~\bibnamefont {Liu}}, \bibinfo {author} {\bibfnamefont {A.}~\bibnamefont {Bylinskii}}, \bibinfo {author} {\bibfnamefont {B.}~\bibnamefont {Braverman}}, \bibinfo {author} {\bibfnamefont {F.}~\bibnamefont {Huber}}, \bibinfo {author} {\bibfnamefont {J.}~\bibnamefont {Amato-Grill}}, \bibinfo {author} {\bibfnamefont {A.}~\bibnamefont {Lukin}}, \bibinfo {author} {\bibfnamefont {N.}~\bibnamefont {Gemelke}}, \bibinfo {author} {\bibfnamefont {A.}~\bibnamefont {Keesling}}, \emph {et~al.},\ }\href@noop {} {\bibfield  {journal} {\bibinfo  {journal} {Nature Communications}\ }\textbf {\bibinfo {volume} {16}},\ \bibinfo {pages} {712} (\bibinfo {year} {2025})}\BibitemShut {NoStop}%
\bibitem [{\citenamefont {Soto}\ and\ \citenamefont {Chepiga}(2025)}]{Chepiga2025}%
  \BibitemOpen
  \bibfield  {author} {\bibinfo {author} {\bibfnamefont {J.}~\bibnamefont {Soto}}\ and\ \bibinfo {author} {\bibfnamefont {N.}~\bibnamefont {Chepiga}},\ }\href {https://arxiv.org/abs/2411.05494} {\bibinfo {title} {Numerical investigation of quantum phases and phase transitions in a two-leg ladder of rydberg atoms}} (\bibinfo {year} {2025}),\ \Eprint {https://arxiv.org/abs/2411.05494} {arXiv:2411.05494 [cond-mat.quant-gas]} \BibitemShut {NoStop}%
\bibitem [{SM()}]{SM}%
  \BibitemOpen
  \href@noop {} {\bibinfo {title} {Supplemental material}}\BibitemShut {NoStop}%
\bibitem [{\citenamefont {Cardy}(1984)}]{Cardy1984}%
  \BibitemOpen
  \bibfield  {author} {\bibinfo {author} {\bibfnamefont {J.~L.}\ \bibnamefont {Cardy}},\ }\href {https://doi.org/10.1016/0550-3213(84)90241-4} {\bibfield  {journal} {\bibinfo  {journal} {Nucl. Phys. B}\ }\textbf {\bibinfo {volume} {240}},\ \bibinfo {pages} {514} (\bibinfo {year} {1984})}\BibitemShut {NoStop}%
\bibitem [{\citenamefont {Cardy}(1989)}]{Cardy1989}%
  \BibitemOpen
  \bibfield  {author} {\bibinfo {author} {\bibfnamefont {J.~L.}\ \bibnamefont {Cardy}},\ }\href {https://doi.org/10.1016/0550-3213(89)90521-X} {\bibfield  {journal} {\bibinfo  {journal} {Nucl. Phys. B}\ }\textbf {\bibinfo {volume} {324}},\ \bibinfo {pages} {581} (\bibinfo {year} {1989})}\BibitemShut {NoStop}%
\bibitem [{\citenamefont {Cabra}\ and\ \citenamefont {Na{\'o}}(1994)}]{CABRA_1994}%
  \BibitemOpen
  \bibfield  {author} {\bibinfo {author} {\bibfnamefont {D.}~\bibnamefont {Cabra}}\ and\ \bibinfo {author} {\bibfnamefont {C.}~\bibnamefont {Na{\'o}}},\ }\href {https://doi.org/10.1142/s0217732394001969} {\bibfield  {journal} {\bibinfo  {journal} {Mod. Phys. Lett. A}\ }\textbf {\bibinfo {volume} {09}},\ \bibinfo {pages} {2017} (\bibinfo {year} {1994})}\BibitemShut {NoStop}%
\bibitem [{\citenamefont {Na{\'{o} }n}\ and\ \citenamefont {Trobo}(2011)}]{NT2011}%
  \BibitemOpen
  \bibfield  {author} {\bibinfo {author} {\bibfnamefont {C.}~\bibnamefont {Na{\'{o} }n}}\ and\ \bibinfo {author} {\bibfnamefont {M.}~\bibnamefont {Trobo}},\ }\href {https://doi.org/10.1088/1742-5468/2011/02/p02021} {\bibfield  {journal} {\bibinfo  {journal} {J. Stat. Mech.}\ }\textbf {\bibinfo {volume} {2011}},\ \bibinfo {pages} {P02021} (\bibinfo {year} {2011})}\BibitemShut {NoStop}%
\bibitem [{\citenamefont {Di~Francesco}\ \emph {et~al.}(1997)\citenamefont {Di~Francesco}, \citenamefont {Mathieu},\ and\ \citenamefont {Senechal}}]{yellowbook}%
  \BibitemOpen
  \bibfield  {author} {\bibinfo {author} {\bibfnamefont {P.}~\bibnamefont {Di~Francesco}}, \bibinfo {author} {\bibfnamefont {P.}~\bibnamefont {Mathieu}},\ and\ \bibinfo {author} {\bibfnamefont {D.}~\bibnamefont {Senechal}},\ }\href@noop {} {\emph {\bibinfo {title} {Conformal Field Theory}}}\ (\bibinfo {year} {1997})\BibitemShut {NoStop}%
\bibitem [{\citenamefont {Fröjdh}\ and\ \citenamefont {Johannesson}(1991)}]{FROJDH1991}%
  \BibitemOpen
  \bibfield  {author} {\bibinfo {author} {\bibfnamefont {P.}~\bibnamefont {Fröjdh}}\ and\ \bibinfo {author} {\bibfnamefont {H.}~\bibnamefont {Johannesson}},\ }\href {https://doi.org/https://doi.org/10.1016/0550-3213(91)90026-T} {\bibfield  {journal} {\bibinfo  {journal} {Nucl. Phys. B}\ }\textbf {\bibinfo {volume} {366}},\ \bibinfo {pages} {429} (\bibinfo {year} {1991})}\BibitemShut {NoStop}%
\bibitem [{\citenamefont {Affleck}(2000)}]{AffleckTCI_2000}%
  \BibitemOpen
  \bibfield  {author} {\bibinfo {author} {\bibfnamefont {I.}~\bibnamefont {Affleck}},\ }\href {https://doi.org/10.1088/0305-4470/33/37/301} {\bibfield  {journal} {\bibinfo  {journal} {Journal of Physics A: Mathematical and General}\ }\textbf {\bibinfo {volume} {33}},\ \bibinfo {pages} {6473–6479} (\bibinfo {year} {2000})}\BibitemShut {NoStop}%
\bibitem [{\citenamefont {Hauschild}\ and\ \citenamefont {Pollmann}(2018)}]{tenpy}%
  \BibitemOpen
  \bibfield  {author} {\bibinfo {author} {\bibfnamefont {J.}~\bibnamefont {Hauschild}}\ and\ \bibinfo {author} {\bibfnamefont {F.}~\bibnamefont {Pollmann}},\ }\href {https://doi.org/10.21468/SciPostPhysLectNotes.5} {\bibfield  {journal} {\bibinfo  {journal} {SciPost Phys. Lect. Notes}\ ,\ \bibinfo {pages} {5}} (\bibinfo {year} {2018})},\ \bibinfo {note} {code available from \url{https://github.com/tenpy/tenpy}},\ \Eprint {https://arxiv.org/abs/1805.00055} {arXiv:1805.00055} \BibitemShut {NoStop}%
\end{thebibliography}
%

\appendix 
\onecolumngrid
\pagebreak[4]
\refstepcounter{section}
\section*{Appendix \Alph{section}: Post-measurement $\sigma$ and $\varepsilon$ dictionary}\label{app:A}
\setcounter{equation}{0}
\renewcommand{\theequation}{A\arabic{equation}}
\setcounter{figure}{0}
\renewcommand{\thefigure}{A\arabic{figure}}
The post-measurement states we consider (characterized by post-selected outcomes ${\bf n}$) always exhibit simple unit cells that one can organize as a set of measured sites followed by a set of unmeasured sets.  For all such states the microscopic operator 
\begin{align}
  \displaystyle \hat{\sigma}_\ell^\mathbf{n} \equiv \frac{1}{|\Lambda_\ell^\mathbf{n}|} \sum\limits_{k \in \Lambda_\ell^\mathbf{n}} (-1)^k \hat{n}_k\sim \sigma + \cdots,
  \label{eq:postmeasdict_sig}
\end{align}
with $\Lambda_\ell^{\bf n}$ the set of unmeasured sites within a unit cell labeled by the `center of mass' position $\ell$,
yields the CFT field $\sigma$ as the leading long-distance contribution.  As an example, for $\mathbf{n} = \{n_{5j}=0\}$ the $\Lambda_\ell^{\bf n}$'s take the form
\begin{equation*}
\begin{tikzpicture}
\node at (0,0) {\includegraphics[scale=0.4]{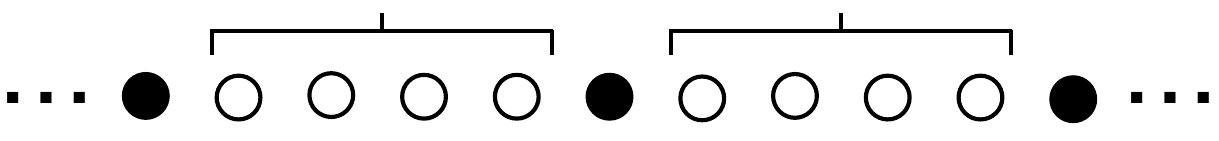}};
\pgfmathsetmacro{\yshift}{-.93}
\node[anchor=west] at (-2.15,1.6 + \yshift) {$\Lambda_{j-5/2}^{\mathbf{n}}$};
\node[anchor=west] at (.98,1.6+ \yshift) {$\Lambda_{j+5/2}^{\mathbf{n}}$};
\node[anchor=west] at (-3.53,0.4-0.03+ \yshift) {$\scriptstyle j-5$};
\node[anchor=west] at (-0.2,0.4-0.03+ \yshift) {$\scriptstyle j$};
\node[anchor=west] at (2.78,0.4-0.03+ \yshift) {$\scriptstyle j+5$};
\end{tikzpicture}
\end{equation*}
While certainly not unique, the mapping in Eq.~\eqref{eq:postmeasdict_sig} democratically incorporates all unmeasured sites within a unit cell into the order-parameter operator.  

When the post-measurement state lacks bond reflection symmetry $R_x$, the $\varepsilon$ field generically appears as the next-leading term represented by the ellipsis in Eq.~\eqref{eq:postmeasdict_sig}.  Preservation of $R_x$, however, allows one to cancel off the $\varepsilon$ contribution---leaving an even less relevant next-leading term and thus streamlining extraction of properties related to $\sigma$.  If the bond-reflection center resides within the set of unmeasured sites (as is the case in the $\mathbf{n} = \{n_{5j}=0\}$ example), then the above $\Lambda_\ell^{\bf n}$ prescription automatically removes the $\varepsilon$ contribution since $\hat{\sigma}^{\bf n}_\ell$ is odd under $R_x$.  For $\mathbf{n} = \{n_{3j},n_{3j+1}=0\}$ where the reflection center resides between consecutive measured sites, our usual $\Lambda_\ell^{\bf n}$ choice leaves an oscillatory $(-1)^\ell \varepsilon$ contribution to $\hat{\sigma}_\ell^{\bf n}$.  For this special case we cancel the oscillatory $\varepsilon$ term by extending $\Lambda_{\ell}^{\bf n}$ to include adjacent unit cells as follows:
\begin{equation*}
\begin{tikzpicture}
\node at (0,0) {\includegraphics[scale=0.4]{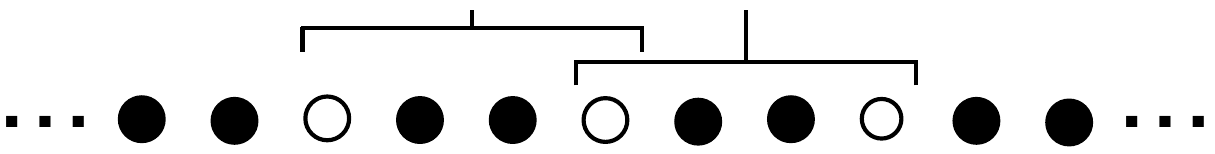}};
\pgfmathsetmacro{\yshift}{1.}
\node[anchor=west] at (-1.5,-0.27+\yshift) {$\Lambda_{j-3/2}^{\mathbf{n}}$};
\node[anchor=west] at (.34,-0.27+\yshift) {$\Lambda_{j+3/2}^{\mathbf{n}}$};
\node[anchor=west] at (-2.28,-1.65+\yshift) {$\scriptstyle j-3$};
\node[anchor=west] at (-0.225,-1.65+\yshift) {$\scriptstyle j$};
\node[anchor=west] at (1.48,-1.65+\yshift) {$\scriptstyle j+3$};
\end{tikzpicture}
\end{equation*}

Outcomes that preserve $R_x$ additionally enable a symmetry-based construction of microscopic operators that yield $\varepsilon$ as the leading contribution (modulo a trivial identity factor).  In particular, we define 
\begin{align}
\hspace{-0.75em}
\displaystyle \hat{\varepsilon}_\ell^\mathbf{n} \equiv \frac{1}{|\Lambda_\ell^\mathbf{n}|} \sum\limits_{k \in \Lambda_\ell^\mathbf{n}} \hat{n}_k \sim c_1^{\bf n} + c_2^{\bf n}\varepsilon + \cdots \quad \text{($R_x$-symmetric } \mathbf{n}\text{)}
\label{eq:postmeasdict_eps}
\end{align}
using exactly the same $\Lambda_{\ell}^{\bf n}$ choices used for $\hat{\sigma}_\ell^{\bf n}$, which precludes a $\sigma$ contribution to the right side by $R_x$ symmetry.  For many post-selected outcomes, even when $R_x$ is broken, one could in principle isolate $\varepsilon$ by searching for a fine-tuned combination of $\hat{n}_k$ operators that cancels off $\sigma$, but we will not pursue such an extraction here.

\refstepcounter{section}
\section*{Appendix \Alph{section}: Intermediate Fixed Point for TCI}
\label{app:B}
\setcounter{equation}{0}
\renewcommand{\theequation}{B\arabic{equation}}
\setcounter{figure}{0}
\renewcommand{\thefigure}{B\arabic{figure}}
As discussed in the main text, $\sigma$ and $\varepsilon$ in Eq.~\eqref{eq:microscopicdla} are both relevant at the TCI point and thus compete nontrivially when the symmetry of the measurement outcome permits both terms.  Here we explore a possible Rydberg array realization of the transition region where these terms compete to a draw---driving a flow to an intervening unstable fixed point.  
To this end, we study generalized post-measurement states $|\psi(\beta,\theta)\rangle = e^{-\frac{\beta}{2}\hat{\mathcal{H}}(\theta)}|\psi_c\rangle$ with $\ket{\psi_c}$ the TCI ground state and 
\begin{align} \hat{\mathcal{H}}(\theta) = \sum\limits_j \left[(-1)^j\sin(\theta)+\cos(\theta)\right]\hat{n}_j. 
\label{eq:generalized measurement}
\end{align} 
At $\theta = 0$, the measurement drives a flow to the $\varepsilon$-driven fixed point for any finite $\beta>0$, since the translation symmetry exhibited by $\hat{\mathcal{H}}$ disallows $\sigma$ from the action.  At $\theta = \pi/2$ by contrast, the measurement promotes CDW configurations by enhancing amplitudes with $n_{2j} = 0$ and $n_{2j+1} = 1$---naturally yielding the $\sigma$-driven fixed point.  Sweeping $\theta$ between these extremes may then cross the intermediate fixed point at some $\beta$-dependent critical angle $\theta_c$.  

\begin{figure}[ht]
\begin{tikzpicture}
  \node[anchor=south west,inner sep=0] (image) at (0,0) {\includegraphics[width=0.75\linewidth]{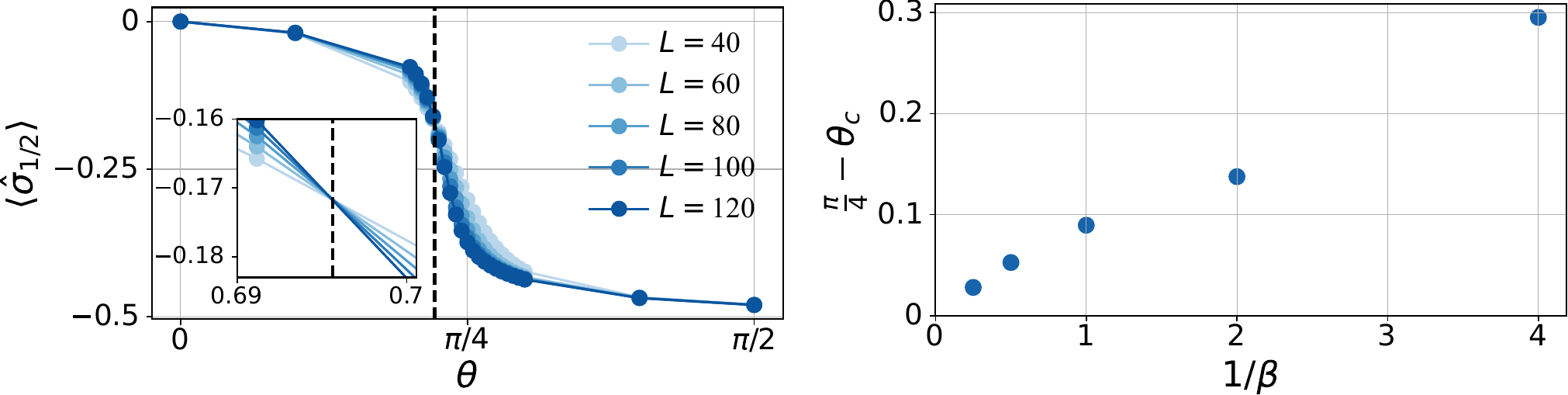}};
  \begin{scope}[x={(image.south east)},y={(image.north west)}]
    \node at (0.05,.99) {\textbf{(a)}};
    \node at (0.534,.99) {\textbf{(b)}};
  \end{scope}
\end{tikzpicture}
\caption{ 
\textbf{Evidence for measurement-induced intermediate fixed point.} (a) Expectation value $\langle\hat{\sigma}_{1/2}\rangle$ versus $\theta$ at $\beta = 1$ for various system sizes.   \emph{Inset:} Zoom-in revealing a scale-invariant point at $\theta_c \approx 0.221\pi$ (vertical dashed line) indicating the unstable intermediate fixed point. 
(b) Extracted $\theta_c$ values at various $\beta$. Axes are chosen to highlight the apparent asymptotic trend, $\theta_c \to \pi/4$ as $\beta\to\infty$.}
\label{fig:findingIFP}
\end{figure}

We confirm this scenario by analyzing the observable $\langle\hat{\sigma}_{1/2}\rangle\equiv \langle\psi(\beta,\theta)|\hat{\sigma}_{1/2}|\psi(\beta,\theta)\rangle$ across various $\beta$ and $\theta$ assuming periodic boundary conditions (hence the results do not depend on the evaluation point).  Figure~\ref{fig:findingIFP}(a) shows $\langle\hat{\sigma}_{1/2}\rangle$ versus $\theta$ at $\beta = 1$, which realizes an increasingly step-like curve as system size $L$ increases.  The inset zoom-in reveals a scale-invariant point at $\theta \approx 0.221\pi$ that we identify as $\theta_c$.  Figure~\ref{fig:findingIFP}(b) illustrates the $\beta$ dependence of $\theta_c$ determined analogously.  

\begin{figure}[ht]
\includegraphics[width=0.5\linewidth]{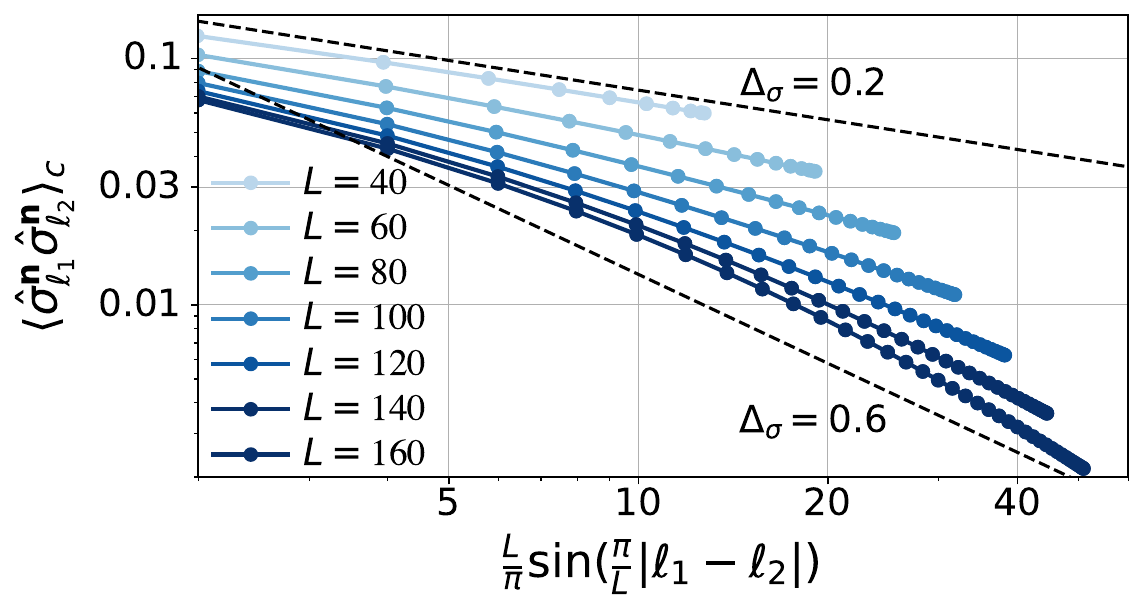}
\caption{Connected order-parameter correlations after applying the projective-measurement protocol with $\mathbf{n}=\{n_{2j} = 0\}$ to TCI ground states on periodic chains.}
\label{fig:IFP_correlators}
\end{figure}

Interestingly, as $\beta \to \infty$, $\theta_c$ appears to asymptotically approach $\pi/4$---which is intuitive from the following perspective.  In the projective-measurement regime $(\beta=\infty)$, $e^{-\frac{\beta}{2}\hat{\mathcal{H}}(\theta)}$ projects the TCI wavefunction onto a symmetric, trivial product state with all $n_j = 0$ for $0\leq \theta< \pi/4$, but instead projects all sites onto a CDW product state for $\pi/4<\theta \leq \pi/2$.  A nontrivially entangled post-measurement state survives uniquely at $\theta = \pi/4$, which intervenes between the adjacent trivial and CDW states (just like the intermediate fixed point expected from the CFT).  Examining Eq.~\eqref{eq:generalized measurement} with $\theta = \pi/4$, one sees that this limit corresponds to our usual projective measurement protocol with $\mathbf{n} = \{n_{2j}=0\}$---suggesting that this relatively high probability outcome may realize (or very nearly realize) the intermediate unstable fixed point with $\Delta_\sigma = 3/5$.  
We further test this possibility by computing the connected order-parameter correlators of the $\{n_{2j}=0\}$ projective post-measurement state.  Despite strong finite-size effects, the results in Fig.~\ref{fig:IFP_correlators} decay much more slowly compared to either the $\sigma$- or $\varepsilon$-driven fixed points (cf.~Fig.~\ref{fig:TCI_FixedPoints}).   
Scaling with system size, however, indicates that the scaling dimension extracted from our simulations exceeds the expected $\Delta_{\sigma} = 3/5$ value as $L$ increases---possibly stemming from imperfect tuning to the TCI point due to finite $V_1$.

\refstepcounter{section}
\section*{Appendix \Alph{section}: Preparation of Critical Ground States}
\label{app:C}
\setcounter{equation}{0}
\renewcommand{\theequation}{C\arabic{equation}}
\setcounter{figure}{0}
\renewcommand{\thefigure}{C\arabic{figure}}
To obtain the critical states used in our numerics, we employ the two-site density matrix renormalization group (DMRG) algorithm to prepare the ground state of the Hamiltonian defined in Eq.~\eqref{eq:model} in the main text. Since our goal is to investigate measurement-altered critical states, it is essential to first accurately prepare the unmeasured critical states within DMRG.

\begin{figure}[ht]
\begin{tikzpicture}
  \node[anchor=south west,inner sep=0] (image) at (0,0) {\includegraphics[width=0.7\linewidth]{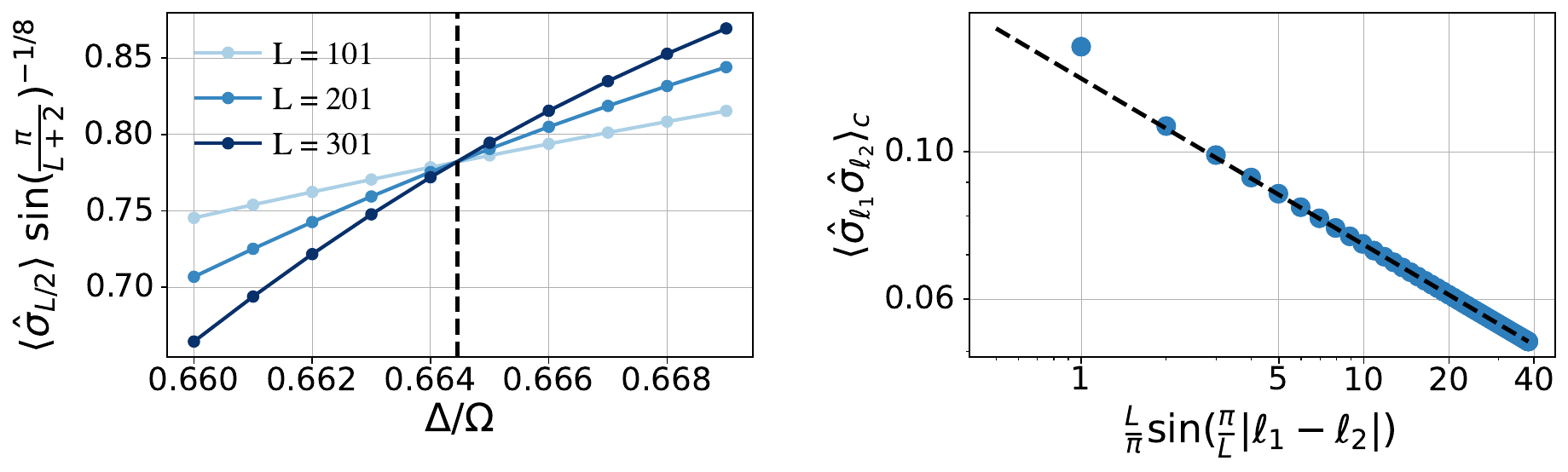}};
  \begin{scope}[x={(image.south east)},y={(image.north west)}]
    \node at (0.072,1.0) {\textbf{(a)}};
    \node at (0.585,1.0) {\textbf{(b)}};
  \end{scope}
\end{tikzpicture}
\caption{Identification of the critical detuning $\Delta_c$ at $V_2=0$ along the Ising critical line. The vertical dashed line in panel (a) marks the value of detuning $\Delta_c = 0.66445$ where $\langle \hat{\sigma}_{L/2} \rangle \cdot \sin(\pi/(L+2))^{-1/8}$ exhibits a curve-crossing for open chains of different size. Panel (b) shows the order parameter connected correlator with respect to the ground state of a Hamiltonian prepared at these extracted critical parameters. The dashed line is the predicted critical scaling for $\Delta_\sigma = 1/8$. }
\label{fig:preparing_crit_ising}
\end{figure}

\begin{figure}[ht]
\begin{tikzpicture}
  \node[anchor=south west,inner sep=0] (image) at (0,0) {\includegraphics[width=0.7\linewidth]{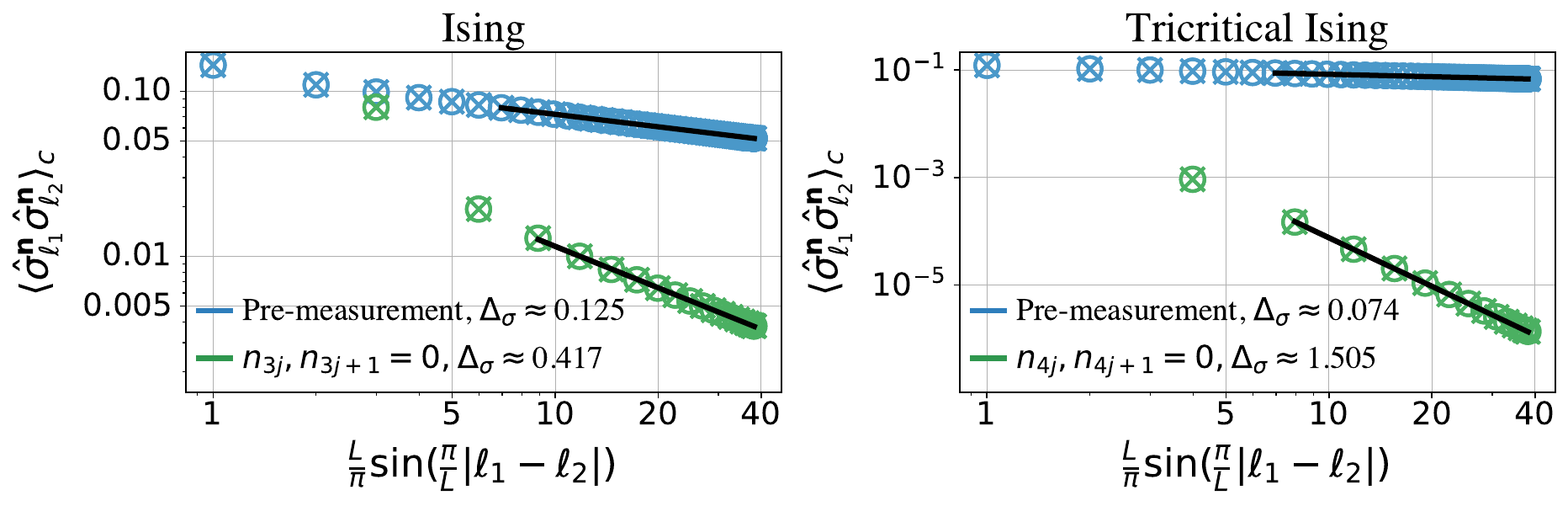}};
  \begin{scope}[x={(image.south east)},y={(image.north west)}]
    \node at (0.06,.95) {\textbf{(a)}};
    \node at (0.54,.95) {\textbf{(b)}};
  \end{scope}
\end{tikzpicture}

\caption{Verification of DMRG convergence in bond dimension. We observe how the observable $\langle\hat{\sigma}^{\mathbf{n}}_{\ell_1}\hat{\sigma}^\mathbf{n}_{\ell_2}\rangle_c$ changes as we double our bond dimension from $\chi = 250$ (denoted by 'o' markers) to $\chi = 500$ (denoted by 'x' markers). We do this for the pre-measurement state (blue symbols), as well as the post-measurement state with the least likely measurement outcome (green symbols). The black line shows a fit of the longest-range $90\%$ of points used to calculate the approximate $\Delta_{\sigma}$'s shown in the legends. Universal predictions are that $\Delta_{\sigma} = 1/8, \ 3/40$ for the pre-measurement states in Ising, TCI respectively.}
\label{fig:bond_dim_conv}
\end{figure}

To realize a Rydberg chain at the Ising critical point, we begin by identifying suitable Hamiltonian parameters along the critical line shown in Fig.~\ref{fig:Setup} in the main text. Following the curve-crossing procedure introduced in Ref.~\onlinecite{Slagle2021}, we sweep the detuning parameter $\Delta$ while fixing the remaining parameters to $\Omega = 1$, $V_1 = 100$ (accounting for the blockaded regime), and $V_2 = 0$. For each value of $\Delta$ and for multiple system sizes with open boundary conditions, we compute the rescaled observable $\langle \hat{\sigma}_{L/2} \rangle \cdot \sin(\pi/(L+2))^{-1/8}$,
which exhibits scale invariance at criticality. More precisely, the critical point is identified as the value of $\Delta=\Delta_c$ where this quantity becomes independent of system size, as shown in Fig.~\ref{fig:preparing_crit_ising}(a). We further verify criticality with periodic boundary chains in Fig.~\ref{fig:preparing_crit_ising}(b) (with additional details provided in Fig.~\ref{fig:bond_dim_conv}(a)) by examining the power-law decay of the connected order parameter correlator at this point in parameter space.

In the case of the TCI point, the critical parameters are known exactly for the closely related integrable model~\cite{FSS} with the Hilbert space constraint $n_j n_{j+1} = 0$ on all sites. These parameters, as provided in Refs.~\onlinecite{FSS,Slagle2021}, are $(\Delta/\Omega)_{\mathrm{TCI}} = -\frac{1}{2}(5\sqrt{5} - 2)$ and $(V_2/\Omega){_\mathrm{TCI}} = -\frac{1}{2}\left(\frac{1 + \sqrt{5}}{2}\right)^{5/2}$. Although we do not explicitly enforce the Hilbert space constraint in our simulations, we approximate it by setting $V_1 = 1000$. Using these parameters places the system close to the TCI point of the constrained model which we verify in Fig.~\ref{fig:bond_dim_conv}(b) (blue data points) showing a scaling of $\langle\hat{\sigma}^{\mathbf{n}}_{\ell_1}\hat{\sigma}^\mathbf{n}_{\ell_2}\rangle_c$ with the (chord) distance $|\ell_2-\ell_1|$ consistent with a power-law dependence, with exponent $\Delta_{\sigma}=3/40$.

 In our DMRG simulations, we terminate the algorithm only when the change in entropy and energy between successive sweeps satisfy $\Delta S < 10^{-5}$ and $\Delta E < 10^{-7}$, respectively. To ensure convergence with respect to the bond dimension $\chi$, we examine the connected correlation function $\langle\hat{\sigma}^{\mathbf{n}}_{\ell_1}\hat{\sigma}^\mathbf{n}_{\ell_2}\rangle_c$ as $\chi$ is doubled (see numerical agreement between "x" and "o" markers in Fig.~\ref{fig:bond_dim_conv}). This analysis is performed for the pre-measurement states in both the TCI and Ising critical points, as well as for the lowest-probability post-measurement states we consider ($\{n_{3j},n_{3j+1}=0\}$ and $\{n_{4j},n_{4j+1}=0\}$ for Ising and TCI, respectively), which we expect to be the most demanding in $\chi$ among those examined.

\refstepcounter{section}
\section*{Appendix \Alph{section}: Extracting Scaling Dimensions}
\label{app:D}
\setcounter{equation}{0}
\renewcommand{\theequation}{D\arabic{equation}}
\setcounter{figure}{0}
\renewcommand{\thefigure}{D\arabic{figure}}
\renewcommand{\thetable}{D\arabic{table}}
Equipped with microscopic mappings that
represent $\sigma$ and (in some cases) $\varepsilon$ after applying a variety of measurement protocols, we
now discuss the extraction of post-measurement scaling dimensions
via DMRG simulations. 

For periodic chains, $\Delta_{\sigma}$ was calculated via a power-law fit of the longest-range $80\%$ of post-measurement connected correlators, $\langle \hat{\sigma}^{\mathbf{n}}_{\ell_1} \hat{\sigma}^{\mathbf{n}}_{\ell_2} \rangle_c$, shown in the insets of Figs.~\ref{fig:Ising_FixedPoints} and~\ref{fig:TCI_FixedPoints} in the main text. For chains with open boundary conditions, translation symmetry is explicitly broken at the edges; $\langle \sigma \rangle$ consequently obtains a spatially dependent expectation value that decays towards zero in the bulk. Reference~\onlinecite{Slagle2021} provides the universal scaling of this expectation value from the boundary of an (unmeasured) odd-length open Rydberg chain:
\begin{align}
\langle \hat{\sigma}_{j+1/2}\rangle \sim \left[\sin\left(\pi\frac{j+1}{L+2}\right)\right]^{-\Delta_{\sigma}}.
\label{eq:slagleOBCfit_main}
\end{align}
In our post-measurement setting, we use the above equation modified by taking 
$j+1/2 \to \ell$ and $\hat\sigma_{j+1/2} \to \hat{\sigma}^{\mathbf{n}}_\ell$.

\begin{figure}[ht]
\begin{tikzpicture}
  \node[anchor=south west,inner sep=0] (image) at (0,0) {\includegraphics[width=0.7\linewidth]{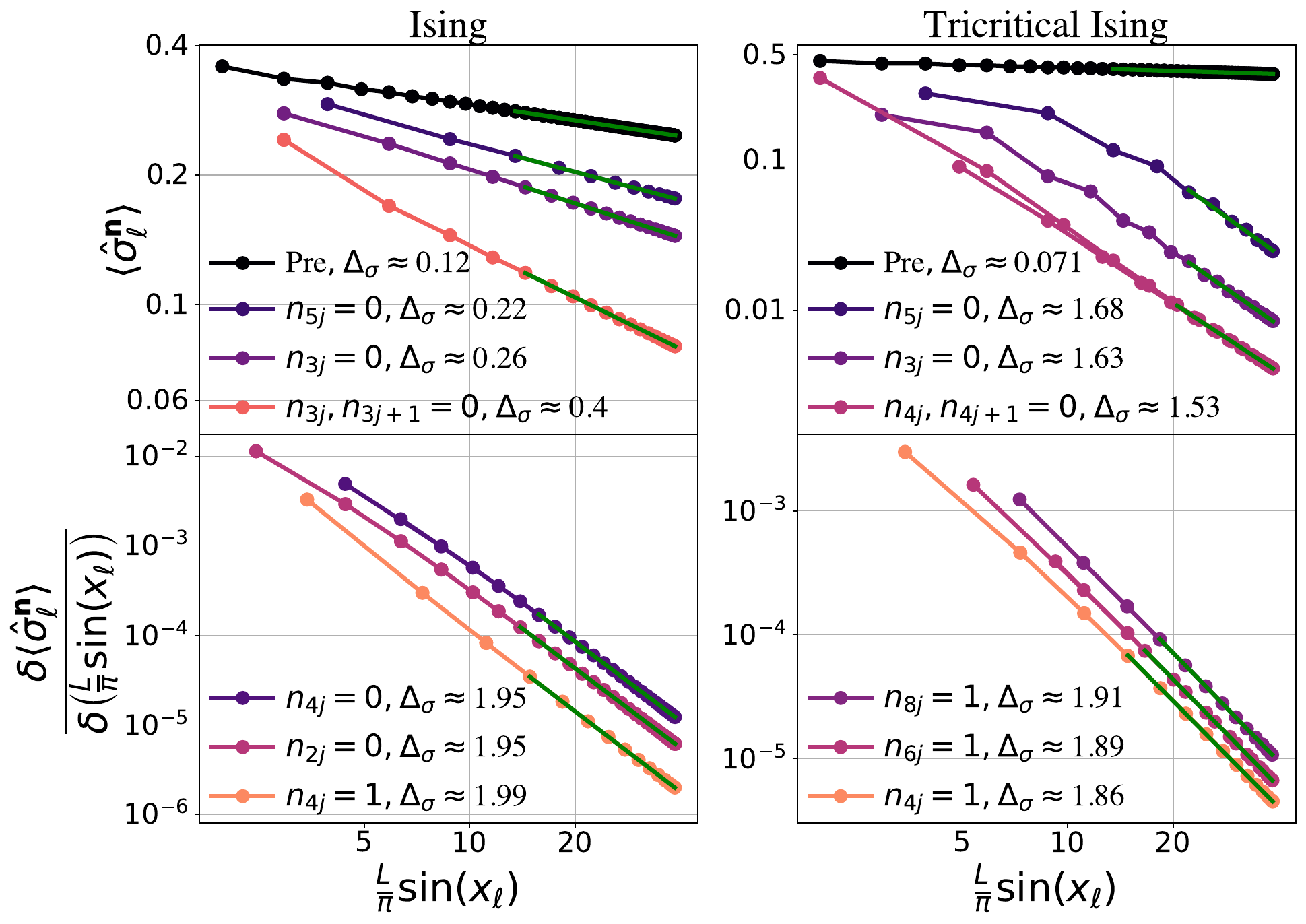}};
  \begin{scope}[x={(image.south east)},y={(image.north west)}]
    \node at (0.185,.86) {\textbf{(a)}};
    \node at (0.64,.86) {\textbf{(b)}};
    \node at (0.185,.475) {\textbf{(c)}};
    \node at (0.64,.475) {\textbf{(d)}};
  \end{scope}
\end{tikzpicture}
\caption{One-point order parameter correlations following measurements of an $L = 121$-site open chain tuned to Ising (left) or tricritical Ising (right) criticality. To give best agreement with boundary CFT predictions, right panels employ a detuning shift, $\Delta\to\Delta-V_2$, on the first and last sites of Eq.~\eqref{eq:model} of the main text, as detailed in Ref.~\onlinecite{Slagle2021}. Data represent measurement protocols yielding $\varepsilon$ (top) and $\sigma$ (bottom) CFT perturbations, with $x_\ell \equiv\pi\frac{\ell+1/2}{L+2}$ on the horizontal axis. Bottom panels show the spatial derivative of the one-point correlator to effectively cancel off a finite $\langle \sigma \rangle$ bulk contribution. The furthest $80\%$ of data from the boundary is used in all cases to fit the corresponding power-law exponent, with the exception of the post-measurement correlators in panel (b) where finite-size effects encourage us to fit only the furthest 2/3 of sites.}
\label{fig:OBC_scaling}
\end{figure}

Figure~\ref{fig:OBC_scaling} examines the open-boundary-condition scaling behavior of the one-point $\sigma$ correlator under various measurement protocols applied to critical Ising and TCI chains. Panel (a) shows $\langle \hat{\sigma}^\mathbf{n}_\ell \rangle$ for $\mathbf{n}$'s that preserve $R_x$ symmetry, corresponding to marginal perturbations to the Ising CFT. As the density of measured sites increases, the numerical results reveal a smooth increase of $\Delta_\sigma$ relative to the pre-measured value of $1/8$.  The extracted $\Delta_\sigma$ values agree well with those obtained from two-point correlators in periodic chains---see Table~\ref{fig:scaling-dim-table}, which provides a summary of the exponents extracted for various measurement outcomes on both open and periodic chains. Panel (b) shows numerical results for $R_x$-symmetric measurement outcomes on the TCI wavefunction generating, in this case, a relevant (symmetric) $\varepsilon$ perturbation. While the analytical prediction is $\Delta_\sigma=3/2$, we observe a slight numerical drift (with the deviation from the predicted $\Delta_{\sigma}$ reaching as large as $12\%$ for the measurement outcome $\{n_{5j}=0\}$) with the measurement strength. We attribute this dependence to finite-size effects (we observe a light improvement with increasing system size, although we are then limited by bond dimension). Moreover, unlike for the $\sigma$-perturbation, the power-law dependence of the correlator for measurement outcomes $n_{5j}=0$ and $n_{3j}=0$, is only observed at sufficiently large distances, preventing us from extracting an accurate power-law exponent.    
Outcomes $\mathbf{n}$ that break $R_x$ symmetry correspond to relevant CFT perturbations that drive a non-zero $\langle \sigma \rangle$ value in the bulk even in the thermodynamic limit. To mod out this bulk contribution and isolate the scaling behavior away from the edges, panels (c) and (d) show the spatial derivative of $\langle \sigma_\ell^{\mathbf{n}} \rangle$ with respect to the horizontal-axis parameterization.   
The fitted slope $\nu$ on the log-log plot yields a scaling dimension $\Delta_\sigma = -\nu - 1$. For both critical Ising and TCI chains, we find $\Delta_\sigma\approx 2$ independent of the measurement protocol---consistent with both analytical predictions and numerical results for periodic chains. 

\begin{table}
  \centering
  \setlength{\tabcolsep}{6pt}     
  \renewcommand{\arraystretch}{1.3} 
  \begin{tabular}{|c|c|c|c|c|c|c|c|}
    \hline
    \multicolumn{8}{|c|}{\bfseries Ising} \\ \hline
    & Pure CFT 
      & \multicolumn{3}{c|}{Relevant Perturbation ($\sigma$)} 
      & \multicolumn{3}{c|}{Marginal Perturbation ($\varepsilon$)} \\ \hline
    & Predic. $\Delta_\sigma=\frac{3}{40}$ 
      & \multicolumn{3}{c|}{Prediction $\Delta_\sigma=2$} 
      & \multicolumn{3}{c|}{Prediction $\Delta_\sigma$ increases with meas. strength} \\ \hline
    $\bf n$ 
      & N/A 
      & $\{n_{4j}=0\}$ 
      & $\{n_{2j}=0\}$ 
      & $\{n_{4j}=1\}$ 
      & $\{n_{5j}=0\}$ 
      & $\{n_{3j}=0\}$ 
      & $\{n_{3j},\,n_{3j+1}=0\}$ 
      \\ \hline
    PBC &
    $0.125233(5)$ & 
    $1.998(5)$&  
    $1.978(3)$&  
    $2.03(2)$&
    $0.2199(1)$&
    $0.26968(4)$ &
    $0.4101(4)$\\ \hline
    OBC & 
    $0.12302(5)$&
    $1.948(3)$&
    $1.952(3)$&
    $1.9921(8)$&
    $0.2165(3)$&
    $0.2639(1)$&
    $0.3978(3)$\\ \hline
  \end{tabular}

  \vspace{1em}

  \begin{tabular}{|c|c|c|c|c|c|c|c|}
    \hline
    \multicolumn{8}{|c|}{\bfseries Tricritical Ising } \\ \hline
    & Pure CFT 
      & \multicolumn{3}{c|}{Relevant Perturbation ($\sigma$)} 
      & \multicolumn{3}{c|}{Relevant Perturbation  ($\varepsilon$)} \\ \hline
    & Predic. $\Delta_\sigma=\frac{1}{8}$ 
      & \multicolumn{3}{c|}{Prediction $\Delta_\sigma=2$} 
      & \multicolumn{3}{c|}{Prediction $\Delta_\sigma=\frac{3}{2}$  }\\ \hline
    $\bf n$ 
      & N/A
      & $\{n_{8j}=1\}$ 
      & $\{n_{6j}=1\}$ 
      & $\{n_{4j}=1\}$ 
      & $\{n_{5j}=0\}$ 
      & $\{n_{3j}=0\}$ 
      & $\{n_{4j},\,n_{4j+1}=0\}$ \\ \hline
    PBC &
    0.074158(8)& 
    1.971(9)&
    1.976(2)&
    1.971(2)&
    1.460(8)&
    1.49(2)&
    1.5101(3)\\ \hline
    OBC &
    0.0711(3)&
    1.914(2)&
    1.887(2)&
    1.856(3)&
    1.68(4)&
    1.63(1)&
    1.529(7)\\ \hline
  \end{tabular}

  \caption{Numerically extracted scaling dimensions $\Delta_{\sigma}$ for Ising and TCI chains in both periodic and open boundary conditions. The uncertainty values (parentheses) are the standard errors from linear regression fits used to obtain $\Delta_{\sigma}$ and are shown as error bars in Figs.~\ref{fig:Ising_FixedPoints} and~\ref{fig:TCI_FixedPoints} from the main text. Note that these uncertainties do not account for finite-size effects or slight deviations from exact critical tuning, both of which we expect contribute to discrepancies from theoretical predictions. }
  \label{fig:scaling-dim-table}
\end{table}

\begin{figure}
\begin{tikzpicture}
\node[anchor=south west,inner sep=0] (image) at (0,0) {\includegraphics[width=0.63\linewidth]{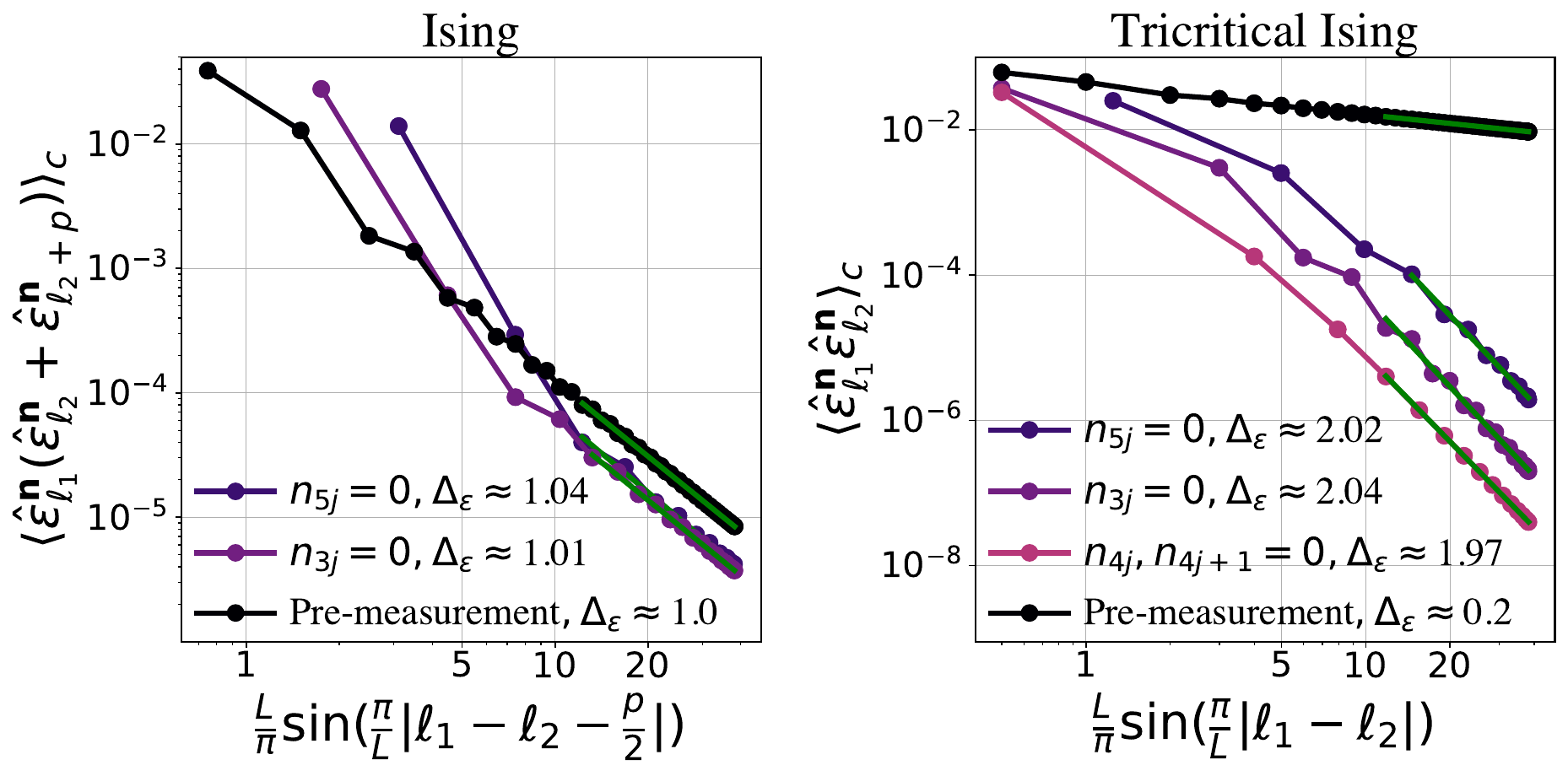}};
  \begin{scope}[x={(image.south east)},y={(image.north west)}]
    \node at (0.08,.92) {\textbf{(a)}};
    \node at (0.58,.92) {\textbf{(b)}};
  \end{scope}
\end{tikzpicture}
\caption{Connected two-point $\varepsilon$ correlators following $R_x$-symmetric measurement protocols in an $L=120$-site periodic critical chain tuned to (a) Ising criticality and (b) the TCI point. 
Figure~\ref{fig:00XEpsilon} analyzes data for the $n_{3j},n_{3j+1}=0$ Ising measurement protocol, which suffers from more severe finite-size effects.  
}
\label{fig:PBC_epseps}
\end{figure}

For $R_x$-preserving measurement outcomes, we can additionally extract the $\varepsilon$ scaling dimension via Eq.~\eqref{eq:postmeasdict_eps} in the main text. 
Figure~\ref{fig:PBC_epseps} demonstrates this extraction through the power-law scaling of the connected correlator $\langle \hat{\varepsilon}_{\ell_1}^\mathbf{n} \hat{\varepsilon}_{\ell_2}^\mathbf{n} \rangle_c$ in a periodic chain prepared at (a) the Ising transition and (b) the TCI point.   
In (a) we perform a local spatial average over adjacent pairs of unit cells to suppress a pronounced oscillatory component, which we attribute to a closely subleading, symmetry-allowed term of the form $(-1)^\ell \partial_x \sigma$ appearing in the ellipsis of Eq.~\eqref{eq:postmeasdict_eps}. In contrast, this averaging is unnecessary for the TCI case in (b), where $\varepsilon$ is significantly more relevant than $\partial_x \sigma$. In both the Ising and TCI case, we find consistency with the CFT predictions of $\Delta_\varepsilon = 1$ and $\Delta_\varepsilon = 2$, respectively.

\subsection{Analysis of $ \langle\varepsilon\varepsilon\rangle_c$ for the $\{n_{3j},n_{3j+1} = 0\}$ outcome}
 
\begin{figure}
\includegraphics[width=0.5\linewidth]{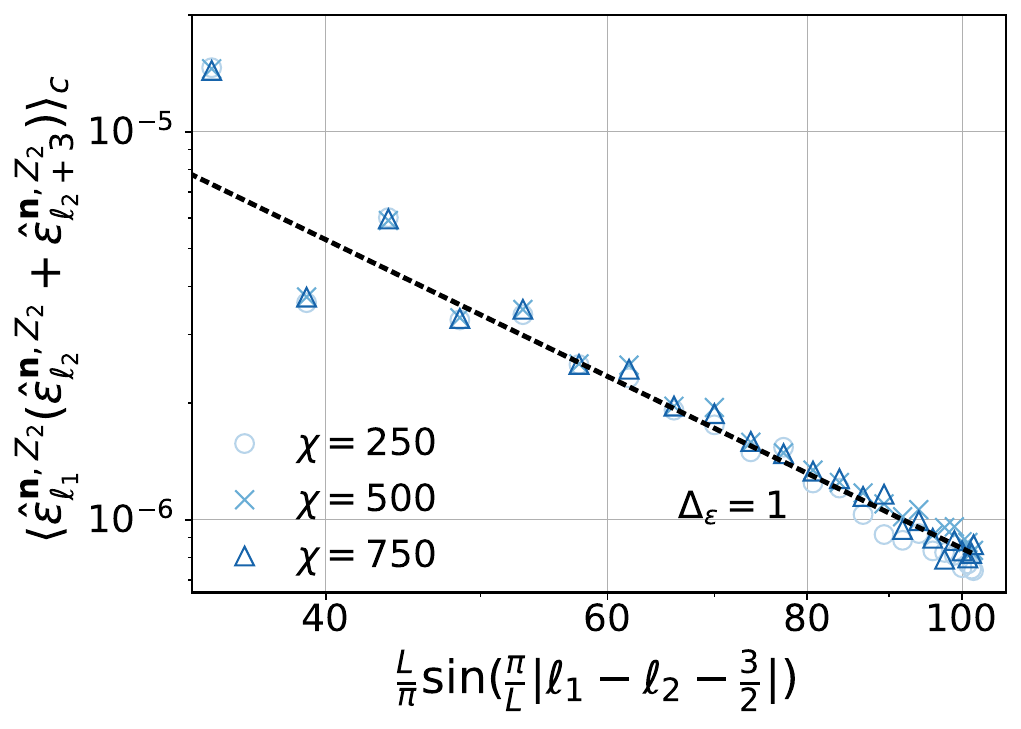}
\caption{Two-unit-cell average of connected correlators of $\hat{\varepsilon}_\ell^{\mathbf{n}, Z_2}$ after applying the $\mathbf{n}=\{n_{3j},n_{3j+1} = 0\}$ measurement protocol on an $L= 180$ Ising critical Rydberg chain with PBC. The dashed line shows the boundary CFT prediction of $\Delta_{\varepsilon} = 1$.} 
\label{fig:00XEpsilon}
\end{figure}

We provide here the analysis of the $\varepsilon$ scaling dimension in the $\{ n_{3j},n_{3j+1} =0\}$ post-measurement state. We employ the following microscopic operator mapping, closely related to the originally prescribed $\varepsilon$ mapping:
\begin{align}
\hat{\varepsilon}_\ell^{\mathbf{n}, Z_2} \equiv \frac{1}{2}\left(\hat{\varepsilon}_{\ell-3/2}^{\mathbf{n}}+\hat{\varepsilon}_{\ell+3/2}^{\mathbf{n}}\right) =  \frac{1}{4}\left(\hat{n}_{\ell-3} + 2\hat{n_\ell}+\hat{n}_{\ell+3}\right)
\end{align}
This choice preserves the symmetries $R_x T_x^3$ and site-reflection $R_x T_x^{-1}$, ensuring that the fields $\sigma$ and $(-1)^{\ell}\partial_x \sigma$, respectively, do not appear in the field-theoretic description. Eliminating the sub-leading field $(-1)^{\ell}\partial_x \sigma$ is crucial due to its pronounced oscillatory effects, which stem from its scaling dimension being very closely sub-leading to that of $\varepsilon$ in the $\{ n_{3j}, n_{3j+1} = 0 \}$ post-measurement setting. We provide evidence of numerical agreement with theoretically predicted scaling in Fig.~\ref{fig:00XEpsilon}.
\end{document}